\documentclass[twocolumn]{aastex631}

\usepackage{amsmath,mathtools}

\newcommand{\hd}{HD\,163296}

\shorttitle{The inner disk rim of \hd{}}
\shortauthors{Chrenko et al.}
\graphicspath{{./}{figures/}}

\begin{document}

\title{The inner disk rim of \hd{}:\\linking radiative
hydrostatic models with infrared interferometry}

\correspondingauthor{Ond\v{r}ej Chrenko}
\email{chrenko@sirrah.troja.mff.cuni.cz}

\author[0000-0001-7215-5026]{Ond\v{r}ej Chrenko}
\affiliation{Charles Univ, Fac Math \& Phys, Astronomical Institute, V Hole\v{s}ovi\v{c}k\'{a}ch 747/2, 180 00 Prague 8, Czech Republic}

\author[0000-0002-9298-3029]{Mario Flock}
\affiliation{Max-Planck-Institut f\"{u}r Astronomie, K\"{o}nigstuhl 17, 69117 Heidelberg, Germany}

\author[0000-0003-4902-222X]{Takahiro Ueda}
\affiliation{Max-Planck-Institut f\"{u}r Astronomie, K\"{o}nigstuhl 17, 69117 Heidelberg, Germany}

\author[0000-0003-2125-0183]{Antoine M\'{e}rand}
\affiliation{European Southern Observatory, Karl-Schwarzschild-Str. 2, 85748 Garching, Germany}


\author[0000-0002-7695-7605]{Myriam Benisty}
\affiliation{Universit\'{e} C\^{o}te d'Azur, Observatoire de la C\^{o}te d'Azur, CNRS, Laboratoire Lagrange, France}
\affiliation{Univ. Grenoble Alpes, CNRS, IPAG, 38000 Grenoble, France}

\author[0000-0002-4327-7857]{Raúl O. Chametla}
\affiliation{Charles Univ, Fac Math \& Phys, Astronomical Institute, V Hole\v{s}ovi\v{c}k\'{a}ch 747/2, 180 00 Prague 8, Czech Republic}



\begin{abstract}

{Previous studies of the protoplanetary 
disk \hd{} revealed that the morphology of its sub-au infrared emission encompasses
the terminal sublimation front of dust grains,
referred to as the inner rim, but also extends into the (supposedly) dust-free region within it.
Here, we present a set of radiative hydrostatic simulations of the inner rim in order to assess how much the rim alone can contribute to the
observed interferometric visibilities $V$, half-light radii $R_{\mathrm{hl}}$, and fractional disk fluxes $\mathcal{F}$ in the wavelength range $1.5$--$13\,\mu\mathrm{m}$.
In our set of models, we regulate the cooling efficiency of the disk via the boundary
condition for radiation diffusion and we also modify the shape of the sublimation front.
We find that when the cooling efficiency is reduced, the infrared photosphere at the rim becomes hotter,
leading to an increase of $R_{\mathrm{hl}}$ sufficient to match the observations.
However, the near-infrared disk flux is typically too low ($\mathcal{F}\simeq0.25$ at $1.5\,\mu\mathrm{m}$), resulting in H-band visibility curves located above the observed data.
We show that the match to the H-band observations up to moderate baselines can be improved when a wall-shaped rather than curved sublimation front is considered. Nevertheless, our model visibilities
always exhibit a bounce at long baselines,
which is not observed, confirming the need for additional emission interior to the rim.
In summary, our study illustrates how the temperature structure and geometry of the inner rim
needs to change in order to boost the rim's infrared emission.
}

\end{abstract}


\section{Introduction}
\label{sec:intro}

Sub-au regions of protoplanetary disks represent the environment that has shaped
precursors of terrestrial planets as well as the numerous population of short-period exoplanets
\citep[e.g.][]{Mulders_etal_2018AJ....156...24M,Petigura_etal_2018AJ....155...89P}
during their early evolution.
{For instance, the transition between the outer dead zone and the inner zone of active 
magnetorotational (MRI) turbulence (at $\approx$$900\,\mathrm{K}$)}
is considered a sweet spot for accumulation of dust grains {\citep[e.g.][]{Varniere_Tagger_2006A&A...446L..13V,Dzyurkevich_etal_2010A&A...515A..70D,Ueda_etal_2019ApJ...871...10U,Jankovic_etal_2022MNRAS.509.5974J}}
as well as a migration trap for planets \citep[e.g.][]{Masset_etal_2006ApJ...642..478M,Flock_etal_2019A&A...630A.147F},
although {the local suppression of the dust drift and planet migration}
seems to strongly depend on disk properties
{\citep[e.g.][]{Schobert_etal_2019ApJ...881...56S,Jankovic_etal_2021MNRAS.504..280J,Chrenko_etal_2022A&A...666A..63C}.}

{One of the possibilities to study sub-au disk regions using observations
lies in the emission of the terminal sublimation front of dust grains, 
hereinafter referred to as the inner disk rim \citep{Dullemond_Monnier_2010ARA&A..48..205D}.}
As {the grains at the rim} equilibrate close to their sublimation temperature, 
{ their thermal emission can become an important contributor to the near-infrared (NIR) excess of} Herbig Ae and Be stars
\citep{Hillenbrand_etal_1992ApJ...397..613H,Lada_Adams_1992ApJ...393..278L,Millan-Gabet_etal_2001ApJ...546..358M,Natta_etal_2001A&A...371..186N}.
{ According to the pioneering models of the inner rim \citep{Dullemond_etal_2001ApJ...560..957D},} no dust grains should exist inwards from the sublimation radius\footnote{{ We point out, however, that large dust grains can cool down efficiently and thus can exist inwards from the conventional sublimation radius \citep{Kama_etal_2009A&A...506.1199K,Klarmann_2018tcl..confE..84K}.}} and the rim { should remain} exposed to irradiation from the central star and heated, { thus becoming} wall-shaped and puffed up. The size of the dust-free region within the rim radius { then} scales with the square root of the stellar luminosity \citep[so-called size-luminosity relation;][]{Monnier_Millan-Gabet_2002ApJ...579..694M}.

{The spatial morphology of the NIR and mid-infrared (MIR) inner-disk emission, accessible 
through the advent of interferometric techniques (with difficulties related
to instrumentation limitations and data sparseness), should in principle trace the inner rim
geometry, manifesting itself via a bright ring or a torus.
While this is sometimes the case and the torus-like emission is indeed observed
\citep{Tuthill_etal_2001Natur.409.1012T,Monnier_Millan-Gabet_2002ApJ...579..694M,Monnier_etal_2005ApJ...624..832M},
there are also cases when an additional emission source located somewhere
between the magnetospheric cavity and the dust sublimation radius is required
\citep[e.g.][]{Eisner_etal_2007ApJ...657..347E,Kraus_etal_2008A&A...489.1157K}.

The emission of inner regions of the protoplanetary disk \hd{}, which is the subject
of this work, is similarly puzzling. \cite{Tannirkulam_etal_2008ApJ...677L..51T}
and \cite{Benisty_etal_2010A&A...511A..74B} found that if only the inner rim torus-like emission is considered,
(i) the visibility curve in H and K bands exhibits a bounce at long baselines inconsistent
with observations, (ii) the observed NIR excess can be recovered only partially,
(iii) and the closure phases typically become too large. By adding a smooth emission source at radii inside the actual
dust rim, they were able to make the visibility curves featureless, increase the NIR flux, and reduce the closure
phase signal. Among possible explanations of the additional emission component
are the optically thin emission of either the hot gas \citep{Tannirkulam_etal_2008ApJ...677L..51T} or refractory dust grains \citep{Vinkovic_etal_2006ApJ...636..348V,Benisty_etal_2010A&A...511A..74B}, but note that a predictive
physical model for neither has ever been put forward.
Further evidence for emission interior to the dust rim was obtained by
\cite{Setterholm_etal_2018ApJ...869..164S} who performed morphological fitting
of the CHARA and VLTI interferometric data to constrain the brightness distribution profile of \hd{}
and concluded that the best-fitting model is a Gaussian-like 2D disk centrally peaked at the star location, without any strong indication
of a sharp dust sublimation radius. The same conclusion was reached
in \cite{Kluska_etal_2020A&A...636A.116K} by means of image reconstruction (but one should
bear in mind the resolution-related issues of image reconstruction on sub-au scales).

In another line of studies, based primarily on fitting prescribed parametric 
brightness distributions directly to the visibility data, 
it was found that the best-fitting model for \hd{} in the VLTI bands H \citep{Lazareff_etal_2017A&A...599A..85L},
K \citep{GRAVITY_2019A&A...632A..53G,GRAVITY_2021A&A...654A..97G}, L, and N \citep{Varga_etal_2021A&A...647A..56V} is a wide ring with an azimuthal modulation.
The azimuthal modulation was found to be evolving with time \citep{Kobus_etal_2020A&A...642A.104K,Varga_etal_2021A&A...647A..56V,GRAVITY_2021A&A...654A..97G},
possibly pointing to a presence of a vortex, a warp, or a variability in the launching zone of the disk wind \citep{Bans_Konigl_2012ApJ...758..100B}.
The resulting width of the emitting ring was again found to be extending within the sublimation radius
of dust grains, re-confirming that the inner rim is not the only
contributor to the NIR and MIR excess in \hd{}.

In summary, it is clear that the inner disk emission of \hd{} can possibly have two components: one arising from the inner rim and one (of unknown origin) from the region inside the rim. However, it remains unclear how the two components compare to one another---is their contribution equally important or is one of them dominant? The question remains unsettled mostly
because the recently used parametric fits \citep{Lazareff_etal_2017A&A...599A..85L,GRAVITY_2021A&A...654A..97G,Varga_etal_2021A&A...647A..56V}
employ only a handful of parameters to avoid degeneracy
and they are also difficult to link directly to physical models.
With this in mind, the strategy of our paper is to start
from a physical model of the inner rim alone \citep[following the framework developed by][]{Flock_etal_2016ApJ...827..144F,Flock_etal_2019A&A...630A.147F}
and see how it compares to the visibility profiles in multiple NIR and MIR bands,
to the half-light radii determined in earlier works, and to previously reported fractional disk fluxes.
Our objective is to answer what it takes to modify the physical model
in order to push some of the synthetic observables closer to the real data.
We mostly focus on modifying the cooling efficiency of the disk
and the shape of the sublimation front.

The aim of our study is by no means to explain the interferometric
observations fully (since we do not model the emission component inside the rim),
nor describe the temporal variability of the inner disk asymmetry (since our model
is by construction static and symmetric). It is rather to set groundwork
for followup studies to help to distinguish how much the inner rim can contribute to
the interferometric signals. In future, our models can be readily combined
with morphological fitting (e.g. by parametrizing an azimuthal asymmetry on top
of one of our base models), or they can help tweaking the relative contribution
between the rim and the interior emission
when a physical description of the latter becomes available.

The manuscript is structured as follows. We describe the radiative hydrostatic method
for deriving the structure of the inner rim in Section~\ref{sec:hydrost_model}.
The list of nominal parameters is given in Section~\ref{sec:models} where we also summarize our
individual models, their boundary conditions, and assumptions for the sublimation temperature of dust grains.
Section~\ref{sec:diagnostics} gives an overview of observables and provides a discussion
of a theoretical link between the interferometric visibilities and half-light radii. 
Our results are presented in Section~\ref{sec:results} and the paper is concluded in Section~\ref{sec:concl}.
Appendix~\ref{sec:app} is devoted to demonstrating the importance of boundary conditions
and the convergence of our models is discussed in Appendix~\ref{sec:app_convergence}.

}

\section{Method}
\label{sec:method}

\subsection{Radiative hydrostatic disk structure}
\label{sec:hydrost_model}

We use the radiative hydrostatic approach of
\cite{Flock_etal_2016ApJ...827..144F,Flock_etal_2019A&A...630A.147F}
to calculate the distribution of the disk gas density $\rho$, dust density $\rho_{\mathrm{d}}$,
and temperature $T$.
Our implementation was done in the \textsc{Fargo3D} code
\citep{Benitez-Llambay_Masset_2016ApJS..223...11B}, extending the work of \cite{Chrenko_Nesvorny_2020A&A...642A.219C}.
The model relies on a decoupling between the timescales of thermal relaxation
(driven mostly by the radiation reprocessing), vertical hydrostatic relaxation (driven by the
propagation of sound waves), and disk accretion (driven by the redistribution of
the angular momentum).

The decoupling makes it possible to proceed iteratively and one iteration
can be summarized as follows:
\begin{enumerate}
    \item Starting with an initial guess of $T$ and keeping it fixed, find $\rho$ in hydrostatic
    equilibrium (see Section~\ref{sec:iter_1_density} for constraints and details).
    \item Perform ten sub-iterations of:
    \begin{enumerate}
        \item Radially integrated optical depths to stellar irradiation $\tau$
        (which depend on the dust-to-gas ratio $f_{\mathrm{d2g}}$ that sets the local 
        optical depth of dust in each grid cell; see Section~\ref{sec:iter_2_opacity} and Equation~\ref{eq:tau}),
        \item Dust-to-gas ratio $f_{\mathrm{d2g}}$ (which has to depend on $\tau$
        in order to properly resolve irradiation absorption; see Section~\ref{sec:iter_3_dust} and Equation~\ref{eq:d2g})
    \end{enumerate}
    \item Keeping $\rho$, $\tau$, and $f_{\mathrm{d2g}}$ fixed, evolve two-temperature energy Equations
    (\ref{eq:e_int}) and (\ref{eq:e_rad}) over a reasonably chosen time step $\mathrm{d}t$ while accounting for the disk heating due to stellar irradiation and viscous heating (Section~\ref{sec:iter_4_temper}).
    \item Return to the beginning with the new temperature field.
\end{enumerate}

\begin{table}[!t]
    \centering
    \caption{Fiducial parameters for the radiative hydrostatic model.}
    \begin{tabular}{ll}
    \hline
    \hline
    Grid size (radial x vertical)  &  $N_{r}\times N_{\phi} = 4096 \times 256$ \\
    Opening angle of the domain & $\Delta\phi=\pm0.24\,\mathrm{rad}$ \\
    Inner radial boundary & $r_{\mathrm{in}}=0.1\,\mathrm{au}$ \\
    Outer radial boundary     & $r_{\mathrm{out}}=15\,\mathrm{au}$ \\
    MRI transition temperature & $T_{\mathrm{MRI}}=900\,\mathrm{K}$ \\
    MRI-active viscosity & $\alpha_{\mathrm{MRI}}=10^{-1}$ \\
    Dead-zone viscosity & $\alpha_{\mathrm{DZ}}=10^{-3}$ \\
    Dust opacity at $T_{\mathrm{s}}$ & $\kappa_{\mathrm{d}}(T_{\mathrm{s}})=751\,\mathrm{cm}^{2}\,\mathrm{g}^{-1}$ \\
    Dust opacity at $T_{\star}$ & $\kappa_{\mathrm{d}}(T_{\star})=1878\,\mathrm{cm}^{2}\,\mathrm{g}^{-1}$ \\
    Gas opacity & $\kappa_{\mathrm{gas}}=10^{-5}\,\mathrm{cm}^{2}\,\mathrm{g}^{-1}$ \\
    Maximum dust-to-gas ratio & $f_{\mathrm{d2g,max}}=10^{-3}$ \\
    Minimum dust-to-gas ratio & $f_{\mathrm{d2g,min}}=10^{-10}$ \\
    Mean molecular weight & $\mu=2.3$ \\
    Adiabatic index & $\gamma=1.43$ \\
    Stellar temperature & $T_{\star}=9000\,\mathrm{K}$ \\
    Stellar radius & $R_{\star}=1.87\,R_{\odot}$ \\
    Stellar mass & $M_{\star}=1.95\,M_{\odot}$ \\
    Stellar luminosity & $L_{\star} \simeq 20.6\,L_{\odot}$ \\
    Mass accretion rate & $\dot{M} = 3.24\times10^{-8} \,M_{\odot}\,\mathrm{yr}^{-1}$ \\
    \hline
    \end{tabular}
    \tablecomments{Stellar parameters of \hd{} are adopted from \cite{Wichittanakom_etal_2020MNRAS.493..234W}
    and the mass accretion rate is from \cite{Fairlamb_etal_2015MNRAS.453..976F}.}
    \label{tab:param}
\end{table}

\begin{table*}[!t]
    \centering
    \caption{Overview of individual models.}
    \begin{tabular}{ll}
    \hline
    \hline
    M1 & Parameters from Table~\ref{tab:param}, {cold boundary for $E_{\mathrm{R}}$ (Section~\ref{sec:models}), $T_{\mathrm{s}}$ given by Equation~(\ref{eq:ts}})\\
    M2 & Parameters from Table~\ref{tab:param}, {warm boundary for $E_{\mathrm{R}}$ (Section~\ref{sec:models}), $T_{\mathrm{s}}$ given by Equation~(\ref{eq:ts}}) \\
    M3 & {Parameters from Table~\ref{tab:param}, warm boundary for $E_{\mathrm{R}}$ (Section~\ref{sec:models}), uniform $T_{\mathrm{s}}=1350\,\mathrm{K}$ } \\
    M3Fe & {as M3, but with a modified dust composition (Section~\ref{sec:M3fe})
    and uniform $T_{\mathrm{s}}=1550\,\mathrm{K}$ } \\
    \hline
    \end{tabular}
    \label{tab:models}
\end{table*}

\subsubsection{Density distribution of gas}
\label{sec:iter_1_density}

\begin{figure}
    \centering
    \includegraphics[width=0.98\columnwidth]{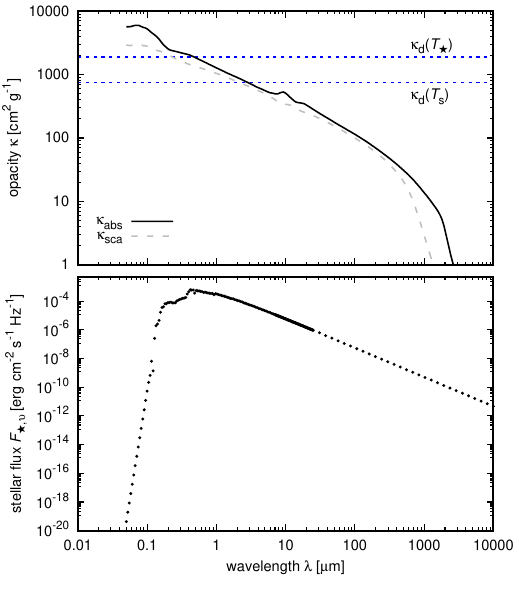}
    \caption{Top: Opacity of dust grains as a function of the wavelength $\lambda$.
    We show the absorption opacity $\kappa_{\mathrm{abs}}$
    (solid black curve) and the scattering opacity $\kappa_{\mathrm{sca}}$
    (dashed grey curve). Dotted blue horizontal lines mark the values
    of Planck-averaged opacities to thermal emission $\kappa_{\mathrm{d}}(T_{\mathrm{s}})$
    and to stellar irradiation $\kappa_{\mathrm{d}}(T_{\star})$.
    Planck-averaged opacities are utilized in our hydrostatic models with radiative diffusion,
    while the wavelength-dependent opacities are used to ray trace synthetic images
    using \textsc{Radmc-3D}. Bottom: Stellar spectrum used in \textsc{Radmc-3D}. 
    Individual points represent the frequency sampling.
    }
    \label{fig:opac}
\end{figure}

At the beginning of each iteration, we fix the temperature
field and solve the
equations of the radial-vertical hydrostatic equilibrium
in spherical coordinates. In a compact form \citep[e.g.][]{Chrenko_Nesvorny_2020A&A...642A.219C},
one can write
\begin{equation}
  \frac{\partial P}{\partial \phi} = r\left( \frac{\partial P}{\partial r} + \rho\frac{GM_{\star}}{r^{2}} \right)\frac{1}{\tan{\phi}} \, ,
  \label{eq:hydrost}
\end{equation}
where $P$ is the thermal pressure, $\phi$ is the colatitude, $r$ is the radius, 
$G$ is the gravitational constant, and $M_{\star}$
is the mass of the central star.
The azimuthal dimension $\theta$ is ignored, assuming an axisymmetric
solution. The radial spacing of the grid is logarithmic and the vertical spacing
is equidistant.

Equation~(\ref{eq:hydrost}) can only be solved along with suitable closure
relations, the first one being the ideal gas equation of state
\begin{equation}
    P = (\gamma-1)\epsilon = (\gamma-1)\rho c_{V} T \, ,
    \label{eq:eos}
\end{equation}
where $\gamma$ is the adiabatic index, $\epsilon$ is the internal energy density of gas,
and $c_{V}$ is the specific heat at constant volume.

Additionally, we assume that the disk is viscously evolving and its
mass accretion rate $\dot{M}$ is uniform. Then the equation
\begin{equation}
    \dot{M} = 3\pi\nu\Sigma \, ,
    \label{eq:mdot}
\end{equation}
where $\nu$ is the effective viscosity, provides a constraint on the gas surface
density $\Sigma$.
The viscosity is parametrized via the \cite{Shakura_Sunyaev_1973A&A....24..337S}
prescription $\nu=\alpha c_{\mathrm{s}}^{2}/\Omega_{\mathrm{K}}$, where $c_{\mathrm{s}}=\sqrt{\gamma P/\rho}$
is the adiabatic sound speed and $\Omega_{\mathrm{K}}$ is the Keplerian angular frequency.
We point out that a density-weighted vertical average
of $\nu$ is used when evaluating Equation~(\ref{eq:mdot}) \citep[see][]{Chrenko_Nesvorny_2020A&A...642A.219C}.
Furthermore, our $\alpha$-parametrization mimics the ionization transition that separates an inner
region where the MRI is active
and an outer dead zone (DZ) where the MRI is suppressed.
Following \cite{Flock_etal_2016ApJ...827..144F}, we write
\begin{equation}
    \alpha(T) = \left(\alpha_{\mathrm{MRI}}-\alpha_{\mathrm{DZ}}\right)\left[\frac{1-\mathrm{tanh}\left(\frac{T_{\mathrm{MRI}}-T}{70\,\mathrm{K}}\right)}{2}\right] + \alpha_{\mathrm{DZ}} \, ,
    \label{eq:alpha}
\end{equation}
where $T_{\mathrm{MRI}}$ is the transition temperature,
$\alpha_{\mathrm{DZ}}$ and $\alpha_{\mathrm{MRI}}$
are $\alpha$-viscosities in the dead and active zones, respectively.
Finally, we assume that $\Sigma$ is related to the volume density
in the midplane $\rho_{\mathrm{mid}}=\Sigma/(\sqrt{2\pi}H)$ as if the disk was vertically isothermal,
with $H=c_{\mathrm{s}}/(\sqrt{\gamma}\Omega_{\mathrm{K}})$ being the pressure scale height
($H$ is evaluated from $c_{\mathrm{s}}$ in the midplane for the purpose of estimating $\rho_{\mathrm{mid}}$).

Equipped with the aforementioned closure relations, it is possible to reconstruct
the radial profile of $\rho_{\mathrm{mid}}(r)$ for a fixed temperature field. This serves
as a starting point for solving Equation~(\ref{eq:hydrost}) and thus finding $\rho(r,\phi)$ throughout the rest of the disk in each iteration \citep[see][]{Flock_etal_2016ApJ...827..144F,Chrenko_Nesvorny_2020A&A...642A.219C}.

\subsubsection{Opacities}
\label{sec:iter_2_opacity}
Before the energy (or temperature) is advanced in our iterative scheme, it is necessary to determine
the opacities in each cell. As in \cite{Flock_etal_2016ApJ...827..144F},
we use a simple three-opacity model and we assume that the Planck ($\kappa_{\mathrm{P}})$ and Rosseland ($\kappa_{\mathrm{R}}$) opacities are the same. We define the gas opacity
$\kappa_{\mathrm{gas}}$, the dust opacity to its own thermal emission $\kappa_{\mathrm{d}}(T_{\mathrm{s}})$,
and the dust opacity to stellar irradiation $\kappa_{\mathrm{d}}(T_{\star})$ (where $T_{\mathrm{s}}$ and $T_{\star}$ are the sublimation and stellar temperatures, respectively).
The optical depth to stellar irradiation is then calculated along radial rays
as
\begin{equation}
    \tau(r) = \tau_{0} + \int\limits_{r_{\mathrm{in}}}^{r}\left(\kappa_{\mathrm{gas}}\rho+\kappa_{\mathrm{d}}(T_{\star})\rho_{\mathrm{d}}\right)\mathrm{d}r' \, ,
    \label{eq:tau}
\end{equation}
where $\tau_{0}$ is the optical depth inwards from our computational grid, at $r<r_{\mathrm{in}}$ \citep[see][]{Flock_etal_2016ApJ...827..144F}. We point out that
$\rho_{\mathrm{d}}$ does not necessarily represent the total dust content but
mainly accounts for small grains, which are the dominant opacity contributors.

As for the actual value of $\kappa_{\mathrm{gas}}=10^{-5}\,\mathrm{cm}^{2}\,\mathrm{g}^{-1}$,
we set it very low in order to maintain the innermost dust-free disk regions optically thin \citep[we refer the reader to appendix B of][]{Flock_etal_2019A&A...630A.147F}. To determine $\kappa_{\mathrm{d}}(T_{\mathrm{s}})$
and $\kappa_{\mathrm{d}}(T_{\star})$, we first calculated wavelength-dependendent dust opacities.
We assumed that the dust grains are composed of $62.5\,\%$  astronomical silicate \citep{Draine_2003ApJ...598.1017D} and $37.5\,\%$ amorphous carbon \citep{Preibisch_etal_1993A&A...279..577P}, having a distribution of physical sizes $f(a)\propto a^{-3.5}$ ranging between $3\times10^{-3}$ and $10^{2}\,\mu\mathrm{m}$.
Using the \textsc{optool} code\footnote{Our wavelength-dependent dust opacities can be reproduced with the following command: \texttt{optool astrosil 0.625 c-p 0.375 -a 0.003 100.0 3.5 -mie}.} \citep{Dominik_etal_optool_2021ascl.soft04010D},
we obtained the wavelength-dependent opacities shown in the top panel of Figure~\ref{fig:opac}. Subsequently,
we calculated $\kappa_{\mathrm{d}}(T_{\mathrm{s}})$ as the Planck opacity at $1400\,\mathrm{K}$
(which is a proxy of the temperature in the dusty disk halo)
and $\kappa_{\mathrm{d}}(T_{\star})$ as the Planck opacity at $9000\,\mathrm{K}$
(which is the effective temperature of the irradiating central star).

The composition and size distribution of dust grains at the inner rim
are largely unconstrained and for simplicity, we chose them in analogy to some of the previous works
\citep{Turner_etal_2014ApJ...780...42T,Flock_etal_2019A&A...630A.147F}. To a limited extent,
we varied the optical dust properties in Section~\ref{sec:M3fe}.

\subsubsection{Density distribution of dust}
\label{sec:iter_3_dust}

We treat the dust grains as passive tracers of the gas and track their volumetric
content using the dust-to-gas ratio $f_{\mathrm{d2g}}(r,\phi)=\rho_{\mathrm{d}}/\rho$
calculated as \citep{Flock_etal_2019A&A...630A.147F}
\begin{equation}
  f_{\mathrm{d2g}} =
  \begin{cases}
    f_{\mathrm{d2g,max}} & T<T_{\mathrm{s}} \land \tau>3 \, ,\\
    \!\begin{aligned}
       & f_{\Delta\tau}\left[\frac{1-\mathrm{tanh}\left(\left(\frac{T-T_{\mathrm{s}}}{100\,\mathrm{K}}\right)^{3}\right)}{2}\right] \\
       & \times\left[\frac{1-\mathrm{tanh}\left(2/3-\tau\right)}{2}\right] 
    \end{aligned}           & \text{otherwise.}
  \end{cases}
  \label{eq:d2g}
\end{equation}
where $f_{\mathrm{d2g,max}}$ is the maximum dust-to-gas ratio and $T_{\mathrm{s}}$ is the 
sublimation temperature of dust grains (see Section~\ref{sec:models}). The value of $f_{\mathrm{d2g,max}}$
is somewhat lower (see Table~\ref{tab:param})
compared to the canonical value of $10^{-2}$ to reflect the fact that the
growth of dust grains depletes the sub-$\mu$m-sized grains \citep{Birnstiel_etal_2012A&A...539A.148B}.
To prevent numerical problems, we also define a floor value $f_{\mathrm{d2g}}>f_{\mathrm{d2g,min}}$.
Additionally, the stability of our method is improved by ramping $f_{\mathrm{d2g,max}}$
from $f_{\mathrm{d2g,min}}$ up to the desired value over the first 25 iterations \citep[similarly to][]{Schobert_etal_2019ApJ...881...56S}.

The term $f_{\Delta\tau}=0.2/(\rho_{\mathrm{d}}\kappa_{\mathrm{d}}\Delta r)$ regulates the maximum
increase of the optical depth to stellar irradiation per one grid cell with the radial length $\Delta r$
and allows to resolve the transition between optically thin and thick regions even with a coarse 
grid spacing \citep[see also][]{Kama_etal_2009A&A...506.1199K}.
We also impose an upper limit $f_{\Delta\tau} \leq f_{\mathrm{d2g,max}}$ to prevent
$f_{\Delta\tau}$ from becoming too large in regions with low $\rho_{\mathrm{d}}$.
Since $\tau$, $f_{\mathrm{d2g}}$ and opacities are mutually dependent through Equations (\ref{eq:tau}) and (\ref{eq:d2g}), we perform 10 sub-iterations within each iteration to evaluate them.

\subsubsection{Evolving the temperature}
\label{sec:iter_4_temper}

To finish one iteration, we search for a new temperature field
corresponding to the hydrostatic distribution of gas and dust.
This is done by integrating
the coupled set of energy equations describing the evolution of $\epsilon$ and the
energy density of thermal radiation field $E_{\mathrm{R}}$ \citep{Dobbs-Dixon_etal_2010ApJ...710.1395D}:
\begin{equation}
  \frac{\partial \epsilon}{\partial t} = - \rho\kappa_{\mathrm{P}}\left[ 4\sigma T^{4} - cE_{\mathrm{R}} \right] + Q_{\mathrm{irr}} + Q_{\mathrm{visc}} \, , \\
  \label{eq:e_int}
\end{equation}
\begin{equation}
  \frac{\partial E_{\mathrm{R}}}{\partial t} + \nabla\cdot\vec{F} = \rho\kappa_{\mathrm{P}}\left[4\sigma T^{4} - cE_{\mathrm{R}}  \right] \, , \\
  \label{eq:e_rad}
\end{equation}
where $\kappa_{P}=\kappa_{\mathrm{gas}}+f_{\mathrm{d2g}}\kappa_{\mathrm{d}}(T_{\mathrm{s}})$, $\sigma$ is the Stefan-Boltzmann constant, $c$ is the speed of light, $Q_{\mathrm{irr}}$ is the irradiation heating rate,
$Q_{\mathrm{visc}}$ is the viscous heating rate,
and $\vec{F}$ is the radiation flux.
The radiative energy is transported using the flux-limited diffusion approximation \citep{Levermoe_Pomraning_1981ApJ...248..321L} with the flux limiter of \cite{Kley_1989A&A...208...98K}.
Equations~(\ref{eq:e_int}) and (\ref{eq:e_rad}) are solved in an implicit form \citep{Bitsch_etal_2013A&A...549A.124B,Chrenko_Lambrechts_2019}
using a simple successive over-relaxation method.
In our iteration scheme, the time step to advance Equations (\ref{eq:e_int}) and (\ref{eq:e_rad}) is $\mathrm{d}t_{1}=10^{5}\,\mathrm{s}$ during the first 100 iterations, followed by 100 iterations with $\mathrm{d}t_{2}=10^{8}\,\mathrm{s}$. We let $f_{\mathrm{d2g}}$ evolve only during the first 100 iterations;
afterwards it remains fixed. The number of iterations and time step sizes are chosen empirically:
$\mathrm{d}t_{1}$ is short enough to avoid convergence problems when the disk is being gradually filled
with dust and $\mathrm{d}t_{2}$ is long enough to bring the most optically thick disk regions
to thermal equilibrium by radiation diffusion.

The heating due to the absorption of stellar photons \citep[e.g.][]{Bitsch_etal_2013A&A...549A.124B,Kolb_etal_2013A&A...559A..80K,Chrenko_Nesvorny_2020A&A...642A.219C} is
\begin{equation}
    Q_{\mathrm{irr}} = \frac{L_{\star}}{4\pi r^{2}}\left(e^{-\tau}-e^{-(\tau+\mathrm{d}\tau)}\right)\frac{S_{\mathrm{cell}}}{V_{\mathrm{cell}}} \, ,
    \label{eq:qirr}
\end{equation}
where $L_{\star}$ is the stellar luminosity, $\mathrm{d}\tau$ is
the increment of the optical depth across a grid cell of interest,
$S_{\mathrm{cell}}$ is the irradiated cross-section of the cell,
and $V_{\mathrm{cell}}$ is its volume.

The viscous heating term is \citep[e.g.][]{DAngelo_Bodenheimer_2013ApJ...778...77D}
\begin{equation}
    Q_{\mathrm{visc}} = \frac{1}{2\nu\rho}\mathcal{T}_{ij}\mathcal{T}^{ij} \, ,
\end{equation}
where $\mathcal{T}_{ij}$ are the components of the viscous stress tensor.
During our first experiments with $Q_{\mathrm{visc}}$, we found that a straightforward implementation
of this term leads to fluctuating (non-converging) solutions due to the coupling with Equation~(\ref{eq:alpha}).
The coupling often results in a feedback loop at spurious locations accross the inner disk rim---if $Q_{\mathrm{visc}}$ manages to locally increase $T$ so that $\nu(\alpha,T)$ starts to increase,
the local surface density starts to drop through $\Sigma=\dot{M}/(3\pi\nu)$, thus changing optical depths
and unbalancing the system from thermal equilibrium. Moreover, the temperature fluctuations also directly affect the dust content via Equation~(\ref{eq:d2g}) \citep[see also][]{Schobert_etal_2019ApJ...881...56S}.
To circumvent the aforementioned issues, we considered uniform $\alpha=\alpha_{\mathrm{DZ}}$ for the purpose
of calculating $Q_{\mathrm{visc}}$ \citep{Schobert_etal_2019ApJ...881...56S}.
Although this leads to an inner inconsistency in our model, we think it is a reasonable first approximation
because $Q_{\mathrm{visc}}$ is calculated correctly in the optically thick regions within the disk interior
and the incorrect solution (with too low $\alpha$)
applies mostly inwards from the disk rim where we expect $Q_{\mathrm{irr}}$
to dominate anyway \citep[see, for instance, figure 9 in][]{Flock_etal_2019A&A...630A.147F}.

\subsection{Individual models}
\label{sec:models}

Our main set of models revolves around modifications
of the disk's cooling efficiency and the shape of the dust sublimation front.
The former is achieved by modifying the boundary condition for the radiation
energy density $E_{\mathrm{R}}$ while the latter is achieved by modifying
the prescription for the sublimation temperature of dust grains $T_{\mathrm{s}}$.
Starting with $E_{\mathrm{R}}$, we
consider two sets of boundary conditions.
The first set is referred to as the cold boundary and it sets \citep[e.g.][]{Schobert_etal_2019ApJ...881...56S}
\begin{equation}
    E_{\mathrm{R}} = (1-\exp(-\tau_{\mathrm{bc}}))a_{\mathrm{R}}T_{\mathrm{bc}}^{4} \, ,
    \label{eq:bc_cold}
\end{equation}
at the inner radial boundary, with $a_{\mathrm{R}}$ being the radiation constant,
$\tau_{\mathrm{bc}}=10^{-2}$, and $T_{\mathrm{bc}}=T_{\mathrm{thin}}$ representing the temperature of optically thin gas.
At the outer radial boundary, we prevent the diffusion of photons.
At the boundaries in colatitude, we set $E_{\mathrm{R}}=a_{\mathrm{R}}(5\,\mathrm{K})^{4}$, assuming
a very low ambient temperature.
The cold boundary is motivated by the fact that Equation~(\ref{eq:e_rad}) describes the evolution of the diffusing field
of photons related to thermal radiation, while the field of irradiating photons is split and treated using an explicit 
absorption in Equation~(\ref{eq:e_int}). The cold boundary therefore allows the diffusing photons to escape freely.

The second set is referred to as the warm boundary
and it assumes
\begin{equation}
    \begin{aligned}
    E_{\mathrm{R}}\left(r,\phi\right) =  & \left(E_{\mathrm{R}}^{\mathrm{thin}}-E_{\mathrm{R}}^{\mathrm{thick}}\right)\left[\frac{1-\tanh{\left(\frac{r-r_{\mathrm{fc}}}{0.1r_{\mathrm{fc}}}\right)}}{2}\right] \\
    & + E_{\mathrm{R}}^{\mathrm{thick}} \, ,
    \end{aligned} 
    \label{eq:bc_hot}
\end{equation}
where $r_{\mathrm{fc}}$ is the radius of full dust condensation at all heights
above the midplane \cite[equation 17 in][]{Ueda_etal_2017ApJ...843...49U}.
To evaluate $E_{\mathrm{R}}^{\mathrm{thin}}$, we use Equation~(\ref{eq:bc_cold}) where $T_{\mathrm{bc}}$
is set to the optically thin temperature of a dusty disk \citep[equation 1 in][]{Ueda_etal_2017ApJ...843...49U}
and $\tau_{\mathrm{bc}}=\min(\tau_{\mathrm{mid}},1)$,
$\tau_{\mathrm{mid}}$ being the optical depth to stellar irradiation in the midplane.
Finally, we use $E_{\mathrm{R}}^{\mathrm{thick}}=a_{\mathrm{R}}T_{\mathrm{thick}}^4$, with $T_{\mathrm{thick}}$ corresponding to the
surface temperature of an optically thick passively
irradiated disk \citep[equations 11--15 in][]{Ueda_etal_2017ApJ...843...49U}.
The warm boundary sets a shallower gradient of $E_{\mathrm{R}}$
at the grid edge in colatitude, thus reducing the cooling efficiency
of the disk.

The purpose and influence of the boundary conditions is further
demonstrated and discussed in Appendix~\ref{sec:app}.
The cold boundary leads to disks with temperature profiles similar to thermal Monte Carlo
simulations. The warm boundary leads to temperature profiles similar to
\cite{Flock_etal_2016ApJ...827..144F}.

Regarding $T_{\mathrm{s}}$, we either consider the dust sublimation temperature of silicate grains
\citep{Pollack_etal_1994ApJ...421..615P,Isella_Natta_2005A&A...438..899I}
\begin{equation}
    T_{\mathrm{s}}=2000\,\mathrm{K}\left(\frac{\rho}{1\,\mathrm{g}\,\mathrm{cm}^{-3}}\right)^{0.0195} \, .
    \label{eq:ts}
\end{equation}
or we set it to a uniform and density-independent value of $T_{\mathrm{s}}=1350\,\mathrm{K}$ (or $1550\,\mathrm{K}$ in Section~\ref{sec:M3fe}).
The purpose of Equation~(\ref{eq:ts}) is to account for the change of sublimation conditions with the height above the midplane, which then leads to a curved inner rim \citep{Kama_etal_2009A&A...506.1199K},
while the purpose of the uniform
sublimation temperature is to produce a wall-shaped rim geometry.

If not specified otherwise, all our models use parameters from Table~\ref{tab:param}.
Differences between individual models are specified in Table~\ref{tab:models}.
Basically, we start from a nominal model M1 with a cold boundary and a curved rim.
Then we go to model M2 by switching to the warm boundary. Keeping the warm boundary,
we change the rim geometry to wall-like in model M3. Results of models 
M1--M3 constitute most of Section~\ref{sec:results}; model M3Fe with a 
different dust composition and a larger sublimation temperature
is discussed in Section~\ref{sec:M3fe}, before concluding the paper.

\subsection{Diagnostics}
\label{sec:diagnostics}

\subsubsection{Synthetic images}
\label{sec:synth_images}

We use the Monte Carlo radiative transfer code \textsc{Radmc-3D}
\citep{Dullemond_etal_2012ascl.soft02015D}
to post-process the results of our hydrostatic modelling.
We use $\rho$, $\rho_{\mathrm{d}}$,
and $T$ obtained with the hydrostatic computations as direct inputs for ray tracing
synthetic images of the inner disk.
Although it is possible to recalculate $T$ in \textsc{Radmc-3D} using the thermal
Monte Carlo method, we do not do so since we verified that the resulting temperature would be
similar to our models with the cold boundary (see also Appendix~\ref{sec:app}).

We use the same computational grid as for the hydrostatic calculations,
thus imposing the axisymmetric approximation and the simplest
isotropic scattering mode. We introduce two species in \textsc{Radmc-3D}.
The first species with the density $\rho$ represents the gas,
for which the absorption opacity at each wavelength is considered uniform, $\kappa_{\lambda}=\kappa_{\mathrm{gas}}$, and the scattering opacity is neglected. The second species with the density $\rho_{\mathrm{d}}=f_{\mathrm{d2g}}\rho$ represents the dust
and its $\kappa_{\lambda}$ is shown in Figure~\ref{fig:opac}.
\textsc{Radmc-3D} offers a possibility to thermalize all species together and we apply this option.
This approximation is incorrect in the optically thin dusty halo of the inner rim
where the dust and gas should be decoupled, however, we apply it for the sake of consistency
because our hydrostatic runs are thermalized as well (we use only one temperature to describe
the gas and dust)\footnote{The thermalization in \textsc{Radmc-3D}, together
with the incorporation of the gas-representing species, ensures
that the temperature profile in the dust-free inner disk and 
at the edges of the optically thin dusty halo remains consistent 
between our hydrostatic calculations and thermal Monte Carlo calculations (see Appendix~\ref{sec:app}).}.

The spectrum of the irradiating star (bottom panel of Figure~\ref{fig:opac}) is adopted from the BOSZ database of stellar atmospheric
models \citep{Meszaros_etal_2012AJ....144..120M,Bohlin_etal_2017AJ....153..234B} and corresponds to
the stellar parameters shown in Table~\ref{tab:param}, along with the surface gravity
$\log{g}=4$ \citep{Wichittanakom_etal_2020MNRAS.493..234W}, and Fe/H = 0.2 \citep{Tilling_etal_2012A&A...538A..20T}.
The wavelength sampling of generated photons, represented by the data points in the bottom panel of Figure~\ref{fig:opac}, covers three log-spaced intervals with 100 samples in $0.05$--$7\,\mu\mathrm{m}$, 100 samples in $7$--$25\,\mu\mathrm{m}$, and 30 samples in $25$--$10^{4}\,\mu\mathrm{m}$.
We use $10^{9}$ photon packages for synthetic image calculations,
while the number of scattering photons amounts to $10^{8}$ (for clarity, we emphasize
that scattering is only considered in calculations with \textsc{Radmc-3D}).
Synthetic images are produced at $\lambda\in(1.5,1.75,2,2.2,2.45,2.8,3.5,4.2,5.5,8,10.5,13)\,\mu\mathrm{m}$, covering NIR and MIR bands of the VLTI instruments, assuming the disk inclination
$i=52^{\circ}$ and position angle $\mathrm{PA}=143^{\circ}$ \citep{Varga_etal_2021A&A...647A..56V}. 
The resolution is $3\times10^{3}$ pixels along the image edge, i.e. $0.1\,\mathrm{mas}$ per pixel.
Second-order ray tracing of \textsc{Radmc-3D} is utilized.

\begin{figure*}
    \centering
    \includegraphics[width=0.98\textwidth]{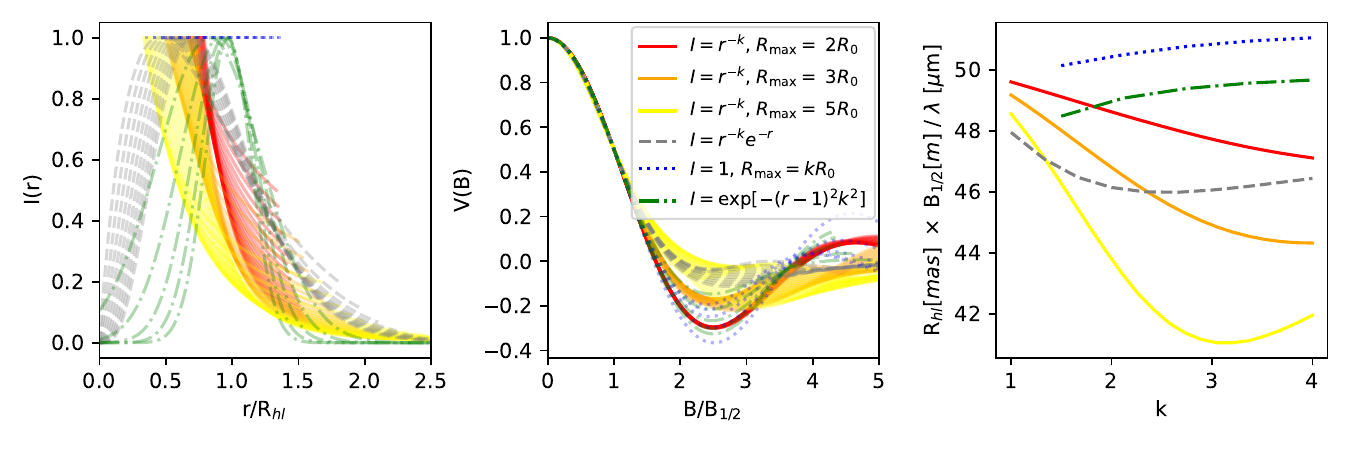}
    \caption{Comparison of $B_\frac{1}{2}\times R_\mathrm{hl}$ for various analytical profiles of $I_\nu$. Each radial profile (left) is plotted as a function of $r/R_\mathrm{hl}$. For each profile, the visibility is computed as a function of $B/B_\frac{1}{2}$ (center). Finally, we compare  (right) $B_\frac{1}{2}\times R_\mathrm{hl}$ as a function of $k$ (which parameterizes the various intensity profiles) and find that it is fairly independent of the function used to represent $I_\nu$.}
    \label{fig:hlr_b12}
\end{figure*}

\subsubsection{Observables}
\label{sec:observables}

Using the synthetic images, we calculate the half-light radii $R_{\mathrm{hl}}$,
fractional disk fluxes $\mathcal{F}=F_{\mathrm{disk}}/F_{\mathrm{tot}}=F_{\mathrm{disk}}/(F_{\mathrm{disk}}+F_{\mathrm{star}})$, and interferometric visibilities $V$ at various $\lambda$ in order to compare them to the real data.
The flux from an individual image pixel is calculated simply by multiplying
the local emission intensity $I_{\nu}$ with the pixel surface area and considering the 
distance of \hd{} being $d=101.5\,\mathrm{pc}$ \citep{Wichittanakom_etal_2020MNRAS.493..234W}.
Then, $F_{\mathrm{disk}}$ is an integral 
over all pixels occupied by our disk model and $F_{\mathrm{tot}}$ is an integral over the entire image.

The half-light radius is defined via \citep{Leinert_etal_2004A&A...423..537L,Varga_etal_2021A&A...647A..56V}
\begin{equation}
    \frac{F_{\mathrm{disk}}}{2} = \int\limits_{0}^{R_{\mathrm{hl}}}2\pi r' I_{\nu}(r')\mathrm{d}r' \, ,
    \label{eq:hlr}
\end{equation}
where $r'$ is the radius from the centre of the image plane and the stellar flux is excluded from the calculation.

The synthetic interferometric visibilities are calculated using codes
\textsc{radmc3dPy}\footnote{Available at \url{https://www.ita.uni-heidelberg.de/~dullemond/software/radmc-3d/manual_rmcpy}}
and \textsc{pmoired} \citep{Merand_2022SPIE12183E..1NM}. 
When analyzing the visibilities, we deproject
the baselines $B$ using \citep[e.g.][]{Tannirkulam_etal_2008ApJ...677L..51T}
\begin{equation}
    B_{\mathrm{eff}} = B \sqrt{\cos{\chi}^{2} + \sin{\chi}^{2}\cos{i}^{2}} \, ,
    \label{eq:B_eff}
\end{equation}
where $i$ is the disk's inclination and
$\chi=\mathrm{PA}_{B} - \mathrm{PA}_{\mathrm{major}}$ is the difference between
the position angle of the given baseline configuration and the major axis of the on-sky disk projection.

\subsubsection{Link between half-light radii and interferometric visibilities}
\label{sec:hlr_vs_vis}

\begin{figure}[!ht]
\centering
    \includegraphics[width=0.98\columnwidth]{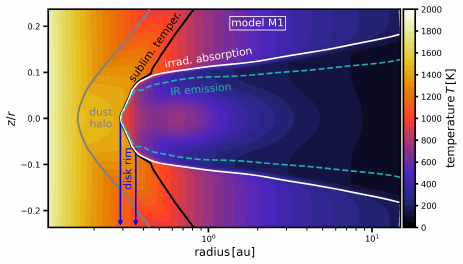}
    \includegraphics[width=0.98\columnwidth]{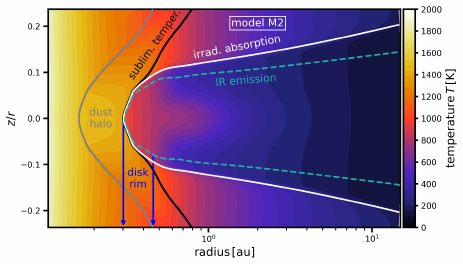}
    \includegraphics[width=0.98\columnwidth]{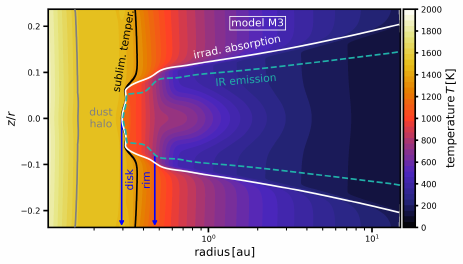}
\caption{Temperature profile (represented by filled contours) in the meridional plane of the disk models M1 (top), M2 (middle), and M3 (bottom).
Individual curves show
where: the dust grains start to condense (solid gray),
the dust is fully condensed
(solid black), the radially-integrated optical depth to stellar irradiation becomes unity
(solid white), and the vertically-integrated optical depth to infrared
emission becomes unity (dashed green). 
Blue arrows delimit the radial range of the inner disk rim.
Since we plot $z/r$ on the vertical axis (where $z$ is the height above the midplane and $r$ is the spherical radius),
we point out that radially propagating
irradiating rays would appear as horizontal lines.}
\label{fig:temper_M1_M2}
\end{figure}

Although the half-light radius is a secondary interferometric observable, we would like
to point out that there is a theoretical argument for a link between $R_{\mathrm{hl}}$ and visibilities
that has not been fully appreciated in prior works. Assuming the disk is viewed face-on\footnote{This discussion is still valid in case of an inclined disk, but the deprojected baseline should be used instead.} and only experiences radial variations in intensity, the interferometric visibility amplitude $V$ is the Hankel transform of the disk profile $I_\nu$ (for the spatial frequency $B/\lambda$), combined linearly with the unit visibility of the unresolved central star:
\begin{equation}
V(B/\lambda) = \frac{\int_0^\infty 2\pi rI_\nu(r)J_0 \left(2\pi r \frac{B}{\lambda}\right) \mathrm{d}r + F_\mathrm{star}}{F_\mathrm{disk} + F_\mathrm{star}} \, .
\label{eq:hankel}
\end{equation}
The Bessel function $J_0$ is an oscillating and vanishing function, so for sufficiently large $B/\lambda$, the disk is fully resolved: the integral in Equation~(\ref{eq:hankel}) vanishes to 0 and the visibility becomes constant as function of the baseline. On the other hand, for very small baselines, the disk is unresolved\footnote{Studies focused on inner disk regions
often assume another contribution of a large-scale over-resolved emission component referred to as halo
\citep[e.g.][]{Lazareff_etal_2017A&A...599A..85L,Setterholm_etal_2018ApJ...869..164S}. Such a component
would result in an addition of $F_{\mathrm{halo}}$ in the denominator of Equation~\ref{eq:hankel}.
It might result in $V\lesssim1$ at very small baselines, which would slightly modify the 
subsequent analysis of this section. However, we neglect this halo component and we also caution
the reader not to confuse it with the optically thin dusty halo defined later in Section~\ref{sec:results}.} ($V\sim1$). So as the baseline increases, the visibility decreases from 1 to the saturation value ${F_\mathrm{star}}/({F_\mathrm{disk} + F_\mathrm{star}})$. For an intermediate spatial frequency $B_\frac{1}{2}/\lambda$, the visibility reaches a mid-point, which can be measured if a sufficiently wide range of baselines length was explored. In that case:
\begin{equation}
\begin{aligned}
    \int_0^{\infty} 2\pi rI_\nu(r)J_0 \left(2\pi r \frac{B_\frac{1}{2}}{\lambda}\right) \mathrm{d}r
    &= 1/2 \int_0^{\infty} 2\pi rI_\nu(r) \mathrm{d}r \\
    &= \frac{F_\mathrm{disk}}{2}  \, ,
\end{aligned}
\label{eq:midvis}
\end{equation}
which is very similar to Equation~(\ref{eq:hlr}) defining the half-light radius, if re-written as:
\begin{equation}
    \int\limits_{0}^{\infty}2\pi r I_{\nu}(r) S(r/R_{\mathrm{hl}})\mathrm{d}r = \frac{F_{\mathrm{disk}}}{2} \, ,
    \label{eq:hlr2}
\end{equation}
with $S(x)=1$ for $x\leq1$ and $S(x)=0$ for $x>1$. Combining the last 2 equations (\ref{eq:midvis} and \ref{eq:hlr2}), we get:
\begin{equation}
    \int\limits_{0}^{\infty} r I_{\nu}(r) J_0 \left(2\pi r \frac{B_\frac{1}{2}}{\lambda}\right)\mathrm{d}r =
    \int\limits_{0}^{\infty}r I_{\nu}(r) S(r/R_{\mathrm{hl}})\mathrm{d}r \, .
\end{equation}

The exact relation between $B_\frac{1}{2}$ and $R_\mathrm{hl}$ in principle depends on the exact profile $I_\nu$. 
To illustrate whether the dependence
is strong or weak, we visualize it in Figure~\ref{fig:hlr_b12}
for a variety of intensity profiles that are typically used to analyse interferometric data. Denoting $I_{0}$ and $R_{0}$ the unit intensity and radius, respectively, we consider power-law
profiles $I(r)=I_{0}r^{-k}$ truncated at $R_{\mathrm{max}}=(2,3,5)\,R_{0}$,
a Poisson-like profile $I(r)=I_{0}r^{-k}e^{-r}$, flat disk profiles
$I(r)=I_{0}$ extending over $r\in(1,k)\,R_{0}$, and Gaussian rings 
$I(r)=I_{0}\exp{(-(r-1)^{2}k^{2})}$.
Parameter $k\in(1,4)$ modulates the shape or extent of individual profiles.
Figure~\ref{fig:hlr_b12} shows where the intensity profiles reach the half-light radius (left panel), where the corresponding visibility
curves reach their mid-point (middle panel),
and how $B_\frac{1}{2}$ relates to $R_\mathrm{hl}$ over the considered range of $k$ (right panel). We find that the dependence on the exact intensity profile is relatively weak,
in the range
\begin{equation}
     R_\mathrm{hl}\,[\mathrm{mas}] = 41..51 \frac{\lambda\,[\mu\mathrm{m}]}{B_\frac{1}{2}\,[\mathrm{m}]} \, .
     \label{eq:range}
\end{equation}

For the sake of clarity, let us emphasize that considerations in this sections
were based on simple parametric radial intensity profiles, while 
our physical disk models generally lead to more complex brightness distributions
(e.g. Section~\ref{sec:synth_im}) with azimuthal variations due to projection
and radiative transfer effects. However, one can also look at previously published
studies to assess whether the exercise with which we obtained
Figure~\ref{fig:hlr_b12} can be generalized. For instance, 
\cite{Varga_etal_2021A&A...647A..56V} showed that the L-band visibility
of \hd{} saturates around 0.1 and $V=0.55$ is reached for $B_\frac{1}{2}[\mathrm{m}]/\lambda[\mu\mathrm{m}]\approx15$. Their best-fit parametric model, which was 
a Gaussian ring with an azimuthal modulation (thus a 2D intensity distribution),
predicts the half-light radius of $3.28\,\mathrm{mas}$, leading to a constant of 49.2 which falls into our range derived in Equation~(\ref{eq:range}).


\section{Results}
\label{sec:results}

\begin{figure}
    \centering
    \includegraphics[width=0.98\columnwidth]{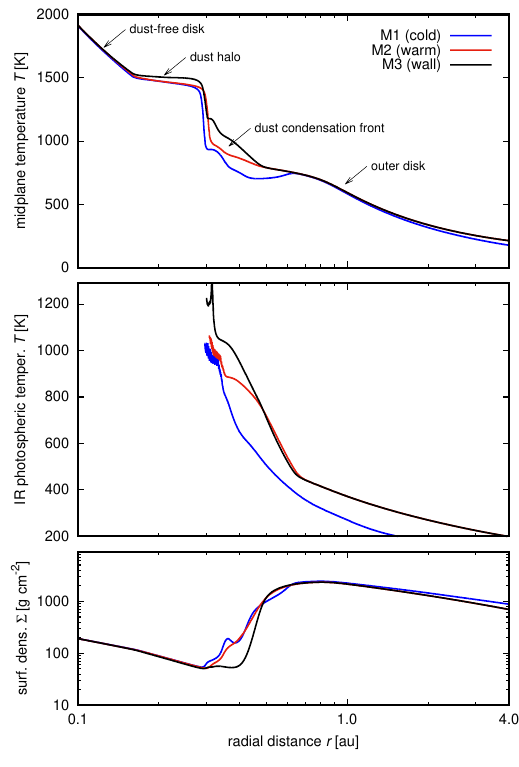}
    \caption{Radial profiles of the midplane
    temperature (top), IR photospheric temperature (middle), and gas surface density $\Sigma$ (bottom) for models M1 (blue curve), M2 (red curve), and M3 (black curve). The top panel distinguishes four characteristic regions of dusty inner rims \citep[see][for details]{Flock_etal_2016ApJ...827..144F,Ueda_etal_2017ApJ...843...49U}.}
    \label{fig:midplane_prof}
\end{figure}

\subsection{Disk structure and temperature profiles}
\label{sec:struct}

Figure~\ref{fig:temper_M1_M2} shows the two-dimensional temperature distribution in
disk models M1, M2, and M3. Additionally, it shows 
several physically distinct surfaces.
The gray curve is the inner boundary of the dust halo
and it shows where the dust grains start to condense in minor quantities \citep{Flock_etal_2016ApJ...827..144F}.
The outer edge of the halo is at the black curve
where $T=T_{\mathrm{s}}$ and the dust becomes fully condensed.
The white curve is the surface where
the optical depth given in Equation~(\ref{eq:tau}) attains unity and
most of the irradiating stellar photons are absorbed.
At the inner rim, the irradiation absorption surface
nearly overlaps with the front of fully condensed dust.
In the outer flaring disk, the irradiation absorption surface delimits 
the disk atmosphere from the disk interior.
The dashed green curve marks the infrared photosphere, i.e. the surface
from which most of the detectable thermal emission originates. This surface, however, is in principle
dependent on the wavelength and the line of sight---for the purpose of Figure~\ref{fig:temper_M1_M2},
we calculated the optical depth to infrared
emission in the direction perpendicular to the midplane (as if the disk was viewed face on) and for the opacity $\kappa_{\mathrm{d}}(T_{\mathrm{s}})$.

Figure~\ref{fig:temper_M1_M2} reveals that the inner disk structure of models M1 and M2
is (unsurprisingly) consistent with general findings of \cite{Kama_etal_2009A&A...506.1199K}
and \cite{Flock_etal_2016ApJ...827..144F}, exhibiting a rounded-off
irradiated inner rim, an optically thin region inwards from the rim, and a flaring disk \citep{Chiang_Goldreich_1997ApJ...490..368C}
outwards from the rim.
Model M3 contains the same regions but its sublimation surface has a nearly vertical wall-like shape,
similar to the classical rim of \cite{Dullemond_etal_2001ApJ...560..957D}. Additionally
the dust halo of model M3 is nearly isothermal.

Figure~\ref{fig:temper_M1_M2} also shows the
radial extent of the inner rim (see the blue arrows). The inner edge
of the rim, $R_{\mathrm{rim}}^{\mathrm{in}}$, 
is defined as the location where the dust fully 
condenses and the irradiation absorption
peaks in the midplane. The outer edge of the rim,
$R_{\mathrm{rim}}^{\mathrm{out}}$, is more difficult to define. \cite{Flock_etal_2016ApJ...827..144F} established
$R_{\mathrm{rim}}^{\mathrm{out}}$ as the local
maximum of the aspect ratio of the infrared photosphere. Our profile of the infrared photosphere, however, has a monotonically 
increasing aspect ratio. Therefore, we define
$R_{\mathrm{rim}}^{\mathrm{out}}$ as the radial
location where the infrared photosphere
has $T=800\,\mathrm{K}$, which is a typical temperature found in \cite{Flock_etal_2016ApJ...827..144F}
at the outer edge of the rim. We emphasize that $R_{\mathrm{rim}}^{\mathrm{in}}$ and
$R_{\mathrm{rim}}^{\mathrm{out}}$ are related to the physical rim size, while 
the characteristic radius of the infrared emission is defined differently (Section~\ref{sec:observables}).

The extent of the inner rim differs between
models M1 and M2; less so between M2 and M3.
We found $R_{\mathrm{rim}}=0.29$--$0.36\,\mathrm{au}$ for M1,
$R_{\mathrm{rim}}=0.3$--$0.46\,\mathrm{au}$ for M2,
and $R_{\mathrm{rim}}=0.3$--$0.47\,\mathrm{au}$ for M3.
Over the extent of the rim, the temperature structure is close to
vertically isothermal, suggesting that stellar irradiation dominates.
Farther out, the temperature along vertical cuts increases towards the midplane
owing to the viscous heating \citep[see also][]{Schobert_etal_2019ApJ...881...56S}.
When moving from model M1 to M2 and then to M3, we see that the temperature over 
the rim extent becomes gradually larger, leading to a warmer and warmer infrared photosphere in this region.
Similarly, the whole interior of the flaring disk region in models M2 and M3
is puffed up. The main cause for the difference between models M1 and M2 is
the boundary condition (Section~\ref{sec:models}, Appendix~\ref{sec:app}), which reduces the cooling efficiency of model M2. The temperature increase found in model M3 is due to the strong frontal
irradiation of the wall-like rim and the radial radiation diffusion. It is important to point out that while the classical wall-like rim of \cite{Dullemond_etal_2001ApJ...560..957D} has a shadowed region right outside
the wall, our model M3 avoids that due to the reduced cooling efficiency combined with viscous heating (see Appendix~\ref{sec:app_convergence} where the role
of viscous heating is further discussed).

Next, Figure~\ref{fig:midplane_prof} compares several characteristic radial profiles of models M1, M2, and M3. Focusing on the midplane temperature first, we can see
that the models differ mostly in the radial range between the dust halo
and the outer disk, $r\simeq0.3$--$0.6\,\mathrm{au}$.
At $r<0.3\,\mathrm{au}$, the disk reaches the optically thin temperature while at $r>0.6\,\mathrm{au}$, the viscous heating
dominates in the midplane, making its temperature independent of the boundary condition.
Despite the relative match in the midplane, however, temperature differences do appear in upper disk layers, as already shown in the context of Figure~\ref{fig:temper_M1_M2}
and highlighted in the middle panel of Figure~\ref{fig:midplane_prof},
which depicts the temperature profile of the IR photosphere. Clearly,
model M3 has the hottest photosphere in the rim region, model M2
is intermediate, and model M1 is the coldest.

\begin{figure}
    \centering
    \includegraphics[width=0.98\columnwidth]{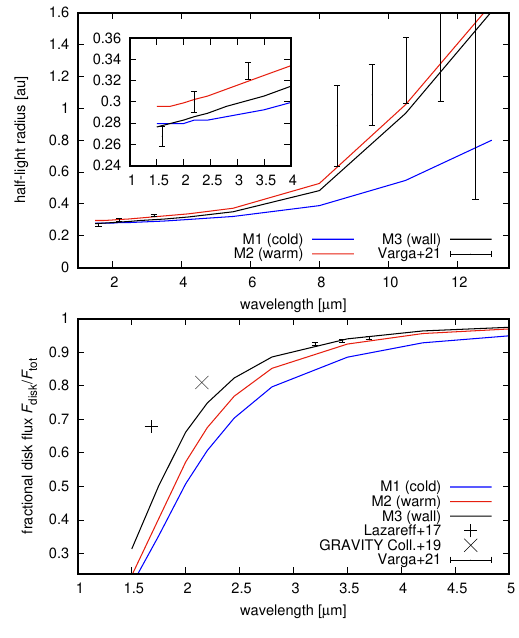}
    \caption{Half-light radius (top) and fractional disk flux (bottom) as functions of the wavelength. We show synthetic
    data corresponding to models M1 (blue), M2 (red), and M3 (black), as well as
    data corresponding to previous morphological studies of the inner disk emission
    \citep[black points and error bars;][]{Lazareff_etal_2017A&A...599A..85L,GRAVITY_2019A&A...632A..53G,Varga_etal_2021A&A...647A..56V}.
    The wavelength range at the bottom panel is limited to $\lambda<5\,\mu\mathrm{m}$ in order
    to highlight the differences at shorter wavelengths (the dependence at larger wavelengths is asymptotic
    and therefore less interesting).
    }
    \label{fig:comparison}
\end{figure}

\subsection{Half-light radii and infrared fluxes}

Figure~\ref{fig:comparison} compares
$R_{\mathrm{hl}}$ and $\mathcal{F}$ (see Section~\ref{sec:observables} for definitions)
derived from our models with observations.
It is necessary to point out that the observational data shown in Figure~\ref{fig:comparison}
are secondary interferometric quantities, in a sense that they are based on parametric brightness
distributions fitted to the interferometric measurements and are therefore model-dependent (cf. Section~\ref{sec:hlr_vs_vis}).
The purpose of the comparison here is simply to get a qualitative understanding of how 
changing various components of our physical model affects the half-light radii and the contribution
of the rim to the overall flux.
Generally, we see that the increase of $R_{\mathrm{hl}}$ and $\mathcal{F}$ 
follows the increase of the photospheric temperature identified between individual models in the previous
Section~\ref{sec:struct}.

Starting with $R_{\mathrm{hl}}$ (top panel in Figure~\ref{fig:comparison}), we can see that
all our models are roughly consistent with previously reported values at NIR wavelengths.
For model M1, however, $R_{\mathrm{hl}}$ increases with $\lambda$ rather weakly and thus 
the half-light radius does not grow enough to match the observations at MIR wavelengths.
Our models M2 and M3, on the other hand, both exhibit a steepening towards the N band and
they seem to match the observations quite well,
with model M2 being slightly nearer the data points.
It seems that the boost of $R_{\mathrm{hl}}$ is mostly driven by 
the warm boundary because both models M2 and M3 use it and their $R_{\mathrm{hl}}$ profiles
are quite similar.

Focusing on $\mathcal{F}$ at $\lambda<5\,\mu\mathrm{m}$ (bottom panel in Figure~\ref{fig:comparison}),
we can see that model M1 has the weakest contribution to the flux.
By adding the warm boundary (in model M2), the disk flux increases, but 
only weakly near $\lambda\sim1.5\,\mu\mathrm{m}$. In model M3,
there is yet another flux increase, most prominent at short wavelengths.
Therefore, the boost of $\mathcal{F}$ at very short wavelengths $\lambda\sim1.5\,\mu\mathrm{m}$
can be achieved when the geometry of the sublimation front becomes wall-shaped
(because models M1 and M2 have rounded rims and their fractional disk flux for $\lambda\sim1.5\,\mu\mathrm{m}$
is nearly the same).

\subsection{Visibilities}

\begin{figure}
    \centering
    \includegraphics[width=0.98\columnwidth]{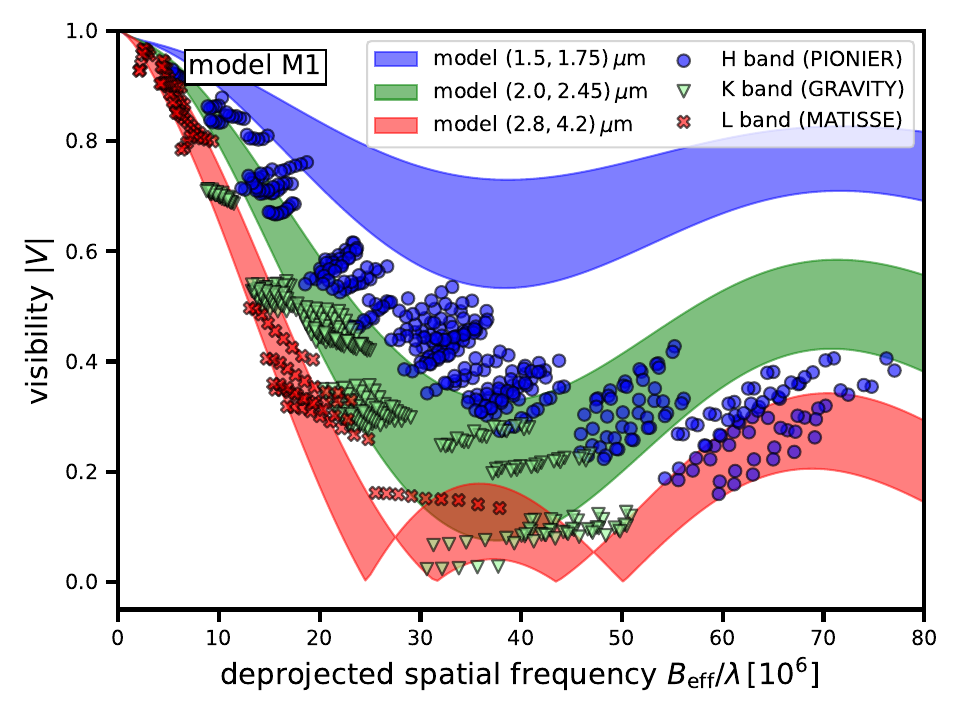}
    \includegraphics[width=0.98\columnwidth]{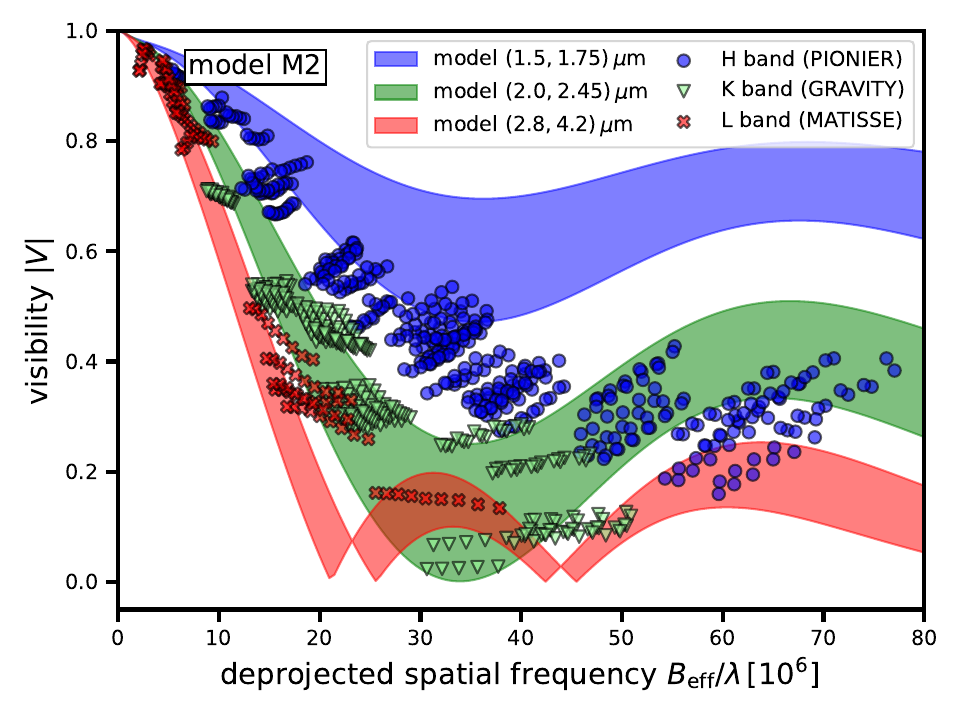}
    \includegraphics[width=0.98\columnwidth]{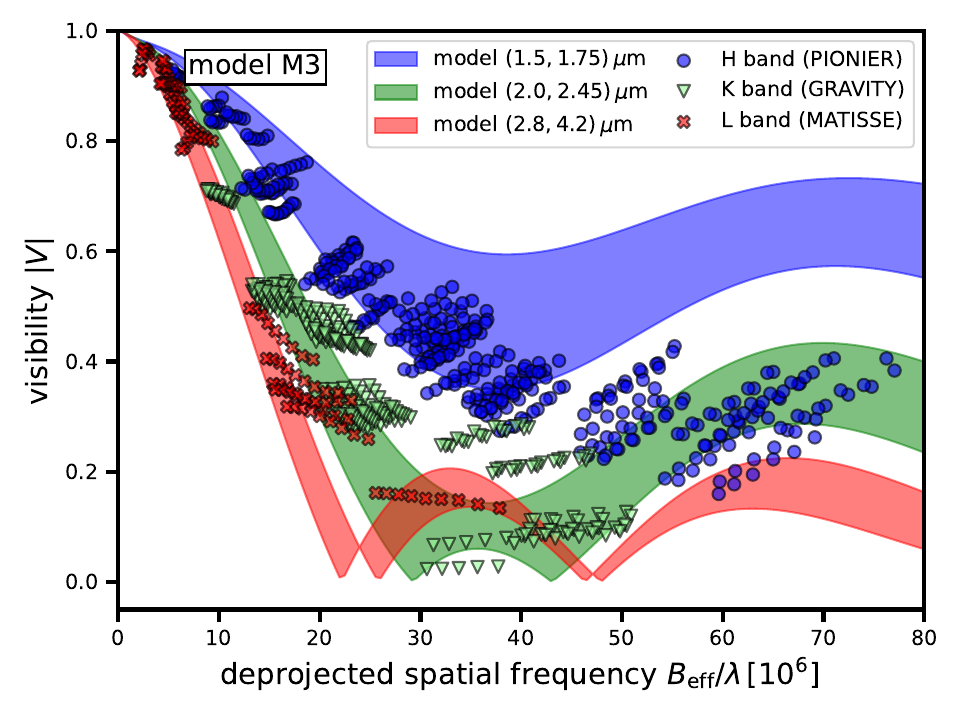}
    \caption{Visibility $|V|$ as a function of the deprojected spatial frequency. 
    Colored areas represent bundles of synthetic visibility curves
    in bands H (blue), K (green), and L (red) 
    that are derived from our models M1 (top), M2 (middle), and M3 (bottom).
    Data points show the VLTI observations from the 2019 epoch taken with
    the PIONIER (circles), GRAVITY (triangles), and MATISSE (crosses) instruments.
    The wavelength intervals of synthetic visibilities and observations are the same and are given in the plot legend.
    }
    \label{fig:vis_L}
\end{figure}

In Figure~\ref{fig:vis_L},
we plot the visibility amplitude $|V|$ as a function of the
deprojected spatial frequency $B_{\mathrm{eff}}/\lambda$ to remove the effect of object inclination.
The data points
show realistic measurements obtained with VLTI in the bands H, K, and L during 2019,
while the colored areas correspond to the visibility profiles of our models,
with the boundary curves calculated at the minimum
(top boundary of each colored area) and maximum (bottom boundary of each colored area) wavelengths of each band.

First, we notice that the mid-point of the visibility curves (between $|V|=1$
and the first bend; see Section~\ref{sec:hlr_vs_vis}) 
at a given band does not seem to strongly depend on the model choice,
which tells us that the half-light radii up to the L band do not differ very much between 
different models.
This is consistent with what is shown in the inset of Figure~\ref{fig:comparison}.

Next, as the disk gets warmer (M1$\rightarrow$M3), the visibility curves start to decay more steeply
and they begin to level off at lower $|V|$. This reflects the fact that
$F_{\mathrm{star}}/F_{\mathrm{tot}}$ decreases as
the fractional disk flux increases\footnote{We refer an interested reader to \cite{Benisty_etal_2010A&A...511A..74B,Dullemond_Monnier_2010ARA&A..48..205D,Lazareff_etal_2017A&A...599A..85L}
where similar interpretations of the rim-induced visibility curves were given.},
as also shown in Figure~\ref{fig:comparison}.
This change is the most prominent in the H band.
The fact that the rim has a torus-like brightness distribution with a sharp edge
(Section~\ref{sec:synth_im}) and that it represents a resolved source leads
to bounces of the visibility curve, especially at long baselines.
While we cannot say much about the presence of absence of bounces in the
displayed K- and L-band observations, they are clearly absent
in the H-band observation, which confirms earlier works
\citep{Benisty_etal_2010A&A...511A..74B,Setterholm_etal_2018ApJ...869..164S}
attributing this mismatch to a presence of a smooth emission source filling the region inside the rim
in \hd{} (e.g. the gas continuum or super-refractory dust species inwards from the sublimation radius).

To provide a simple quantitative comparison between the models and the observations,
we counted the number of observational data points enclosed\footnote{We point out that we take
the observational errors of $V^{2}$ into account.} between the model curves
of each specific band. 
We found that model M1 matches 2\%, 54\%, and 79\% of H-band, K-band, and L-band observations,
respectively. As for model M2, we found an overlap with 15\%, 66\%, and 60\% of H-band, K-band, and L-band observations, respectively. 
Finally, model M3 is consistent with 48\%, 42\%, and 66\% of observations in bands H, K, and L, respectively.
On average, model M3 leads to visibility curves closest to the observations, although 
model M1 is better when focusing on the L band alone and model M2 outperforms the others in the K band.


\begin{figure*}
\centering
\begin{tabular}{cc}
    \includegraphics[width=0.48\textwidth]{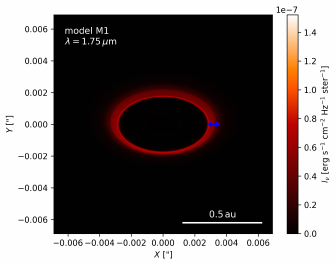} & 
    \includegraphics[width=0.48\textwidth]{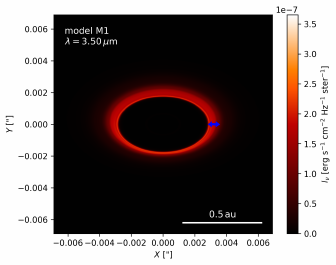} \\
    \includegraphics[width=0.48\textwidth]{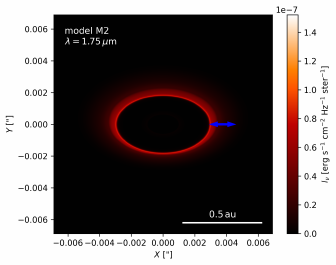} & 
    \includegraphics[width=0.48\textwidth]{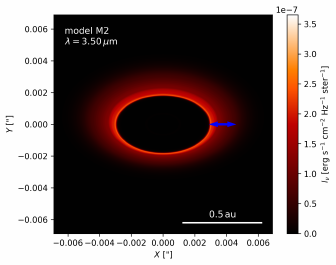} \\
    \includegraphics[width=0.48\textwidth]{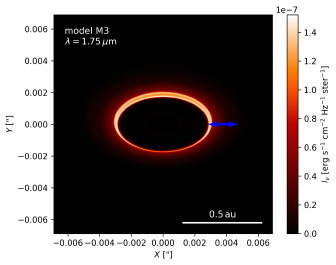} & 
    \includegraphics[width=0.48\textwidth]{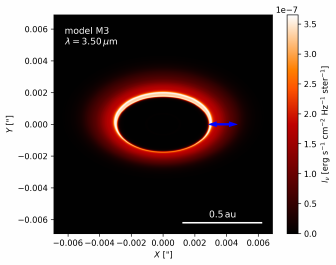} \\
    \end{tabular}
\caption{Synthetic images (emission intensity maps) derived from models M1 (top), M2 (middle), and M3 (bottom) at wavelengths $\lambda=1.75$ and $3.5\,\mu\mathrm{m}$ (left and right column, respectively).
We point out that the intensity is shown in the linear scale and the range of color bars slightly
differs between the left and right column.
The emission of the central star is masked. The 
blue arrow marks the approximate radial range 
of the disk rim defined in Figure~\ref{fig:temper_M1_M2}.
The white scale bar corresponds to $0.5\,\mathrm{au}$.}
\label{fig:image_M1_M2}
\end{figure*}

\subsection{Infrared emission in detail}
\label{sec:synth_im}

Let us now examine the synthetic images\footnote{Synthetic images of our main models M1, M2, M3, and M3Fe are freely available
as FITS files at \url{https://sirrah.troja.mff.cuni.cz/~chrenko/hd163296/}}
themselves (Figure~\ref{fig:image_M1_M2}) and explore the spatial
distribution of infrared emission. Overall, all synthetic images exhibit a dominant torus-like
emission, with the torus being sharply truncated at the inner edge of the rim (the blue arrows can guide the eye as they corresponds to the rim extent shown in Figure~\ref{fig:temper_M1_M2}). 
However, it is important to point out that the width of the
brightest part of the torus is smaller than $1\,\mathrm{mas}$, which is roughly the best 
possible resolution that the VLTI can reach with its most detailed H-band observations. The images
shown here are therefore highly idealized.

Comparing models M1 (top row) and M2 (middle row) first, we can see that their H-band infrared emission
(left column at $\lambda=1.75\,\mu\mathrm{m}$)
is largely similar, despite the differences in the radial range of the rim. This is because 
most of the H-band emission comes from the very tip of the rim, which exhibits a similar grazing angle with respect to
the incoming irradiation (Figure~\ref{fig:temper_M1_M2}) and also a similar
profile of the infrared photosphere at $r\simeq0.3$--$0.35\,\mathrm{au}$ (Figure~\ref{fig:midplane_prof}).
This is consistent with the similarity of the H-band fractional disk flux and visibility curves
of these two models.
On the other hand, the L-band emission (right column at $\lambda=3.5\,\mu\mathrm{m}$)
is more radially extended for model M2, for which it covers roughly the entire radial extent of the rim.
The additional emission of model M2 (dark-red-colored) compared to model M1 is relatively weak but covers a large enough surface area
to increase the L-band flux of model M2 as seen in Figure~\ref{fig:comparison}.
The cause of this difference can be traced back to Figure~\ref{fig:temper_M1_M2} where
the transition from the tip of the inner rim to the flaring outer disk is more abrupt for model M1 than it is for model M2. Model M2, instead, has a noticeable transitional region
that is less exposed to stellar irradiation than the tip of the rim
but more exposed than the outer flaring disk.
Similar geometrical differences can be noticed
in the infrared photosphere of model M2, as well as a temperature bump
between $\simeq$$0.35$--$0.6\,\mathrm{au}$ in the middle panel
of Figure~\ref{fig:midplane_prof}.

Synthetic images of model M3 (bottom row of Figure~\ref{fig:image_M1_M2}) exhibit the largest
absolute intensity compared to models M1 and M2. The bright central part matches the frontally irradiated
wall-like sublimation front, viewed under the inclination angle of the disk. The overall larger intensity,
which also translates to larger fluxes and the previously discussed shifts in the visibility profiles, is yet another manifestation of the hot infrared photosphere.

\subsection{Maximizing the NIR flux}
\label{sec:M3fe}

\begin{figure}
    \centering
    \includegraphics[width=0.98\columnwidth]{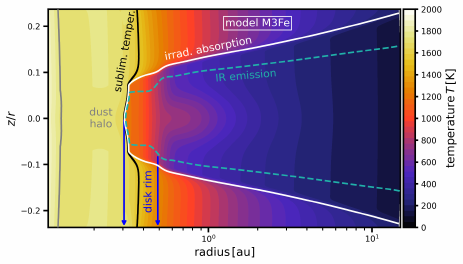}
    \includegraphics[width=0.98\columnwidth]{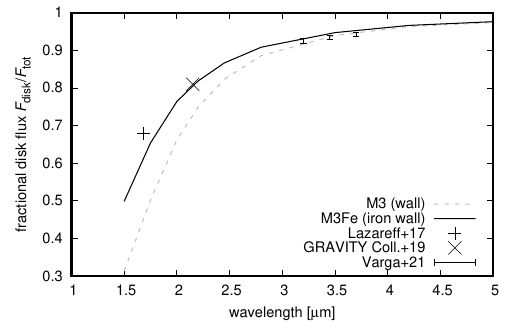}
    \includegraphics[width=0.98\columnwidth]{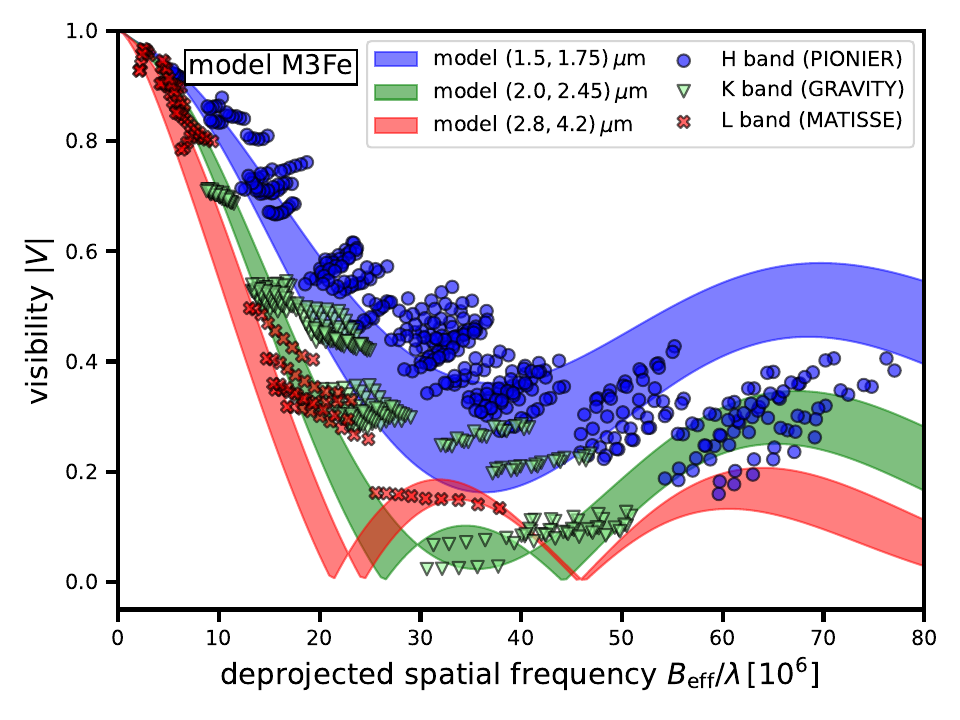}
    \caption{Summary of results for model M3Fe. Top: temperature distribution in the
    meridional plane of the disk (compare with Figure~\ref{fig:temper_M1_M2}).
    Middle: fractional disk flux as a function of the wavelength (compare with Figure~\ref{fig:comparison}).
    Bottom: synthetic visibility curves compared to observations (compare with Figure~\ref{fig:vis_L}).}
    \label{fig:M3Fe}
\end{figure}

Previous findings related to models M1--M3 indicate a close connection between
the NIR flux and the rim temperature. 
In this section, we explore whether it is possible to increase the flux even more by simply considering
a larger sublimation temperature of dust grains, thus making the rim hotter \citep[see also][]{Klarmann_etal_2017A&A...599A..80K}.
To demonstrate this possibility, we computed one additional variation of model M3, designated M3Fe, assuming $T_{\mathrm{s}}=1550\,\mathrm{K}$.

Increasing $T_{\mathrm{s}}$ alone while keeping the other model components fixed
would allow the dust grains to survive closer to the star and the entire disk rim
would shift inwards, making its $R_{\mathrm{hl}}$ inconsistent with the observations.
To keep $R_{\mathrm{hl}}$ comparable to the dependence discussed
in Figure~\ref{fig:comparison}, it is necessary to modify the dust opacities.
We remind the reader that the ratio
\begin{equation}
    \varepsilon = \frac{\kappa_{\mathrm{d}}(T_{\mathrm{s}})}{\kappa_{\mathrm{d}}(T_{\mathrm{\star}})} \, ,
    \label{eq:epsilon}
\end{equation}
determines the optically thin temperature of isolated dust grains
\begin{equation}
    T_{\mathrm{thin}} = \varepsilon^{-1/4}\sqrt{\frac{R_{\star}}{2r}}T_{\star} \, ,
    \label{eq:t_thin}
\end{equation}
as well as the radius where
dust grains condense in the midplane \citep{Monnier_Millan-Gabet_2002ApJ...579..694M,Ueda_etal_2017ApJ...843...49U}:
\begin{equation}
    R_{\mathrm{rim}}^{\mathrm{in}} =
    \frac{1}{2}R_{\star}\epsilon^{-1/2}\left(\frac{T_{\star}}{T_{\mathrm{halo}}}\right)^{2} \, ,
    \label{eq:r_rim_in}
\end{equation}
where $T_{\mathrm{halo}}$ is the temperature
in the halo of the rim.

After testing several dust compositions, we found that $R_{\mathrm{hl}}$ can be preserved
when dust grains composed of solid metallic iron \citep{Henning_Stognienko_1996A&A...311..291H,Woitke_etal_2018A&A...614A...1W}
are considered\footnote{Our wavelength-dependent opacities 
of metallic iron can be reproduced with the \textsc{optool} code as follows: \texttt{optool iron 1.0 -a 0.05 1.0 3.5 -mie}},
with a size
distribution $f(a)\propto a^{-3.5}$ ranging from $0.05$ to $1\,\mu\mathrm{m}$.
Then, $\kappa_{\mathrm{d}}(T_{\mathrm{s}})=873$ and $\kappa_{\mathrm{d}}(T_{\mathrm{\star}})=4187\,\mathrm{cm}^{2}\mathrm{g}^{-1}$, yielding $\varepsilon\simeq1/5$ and $T_{\mathrm{halo}}\simeq1700\,\mathrm{K}$ (the latter was found directly
from our simulation). 
Sublimation temperature of metallic iron sensitively depends on the local chemical conditions
but can in principle reach the assumed value $T_{\mathrm{s}}=1550\,\mathrm{K}$ 
\citep[e.g.][]{Broz_etal_2021NatAs...5..898B}.
We also point out that recent laboratory experiments \citep{Bogdan_etal_2023A&A...670A...6B}
show that metallic iron efficiently and `automatically'
forms from silicates at $T>1200\,\mathrm{K}$,
and thus it can indeed be present at the inner disk rim at large abundances.

The results for model M3Fe are presented in Figure~\ref{fig:M3Fe}.
The temperature distribution in the meridional plane (top panel) is similar to model M3
but the halo and the rim itself are hotter, while the flared outer disk is puffed up even more. 
Looking at the fractional disk flux (middle panel), it is clear that the model now matches observations
even at the shortest H- and K-band wavelengths. As for the visibility trend (bottom panel),
we can see that model M3Fe is a logical continuation of the sequence of panels shown in Figure~\ref{fig:vis_L}:
the visibility curves reflect the increase of the disk flux and so they decrease more steeply.
The H-band synthetic curves are now positioned partially below the set of observations.
The K-band and L-band synthetic curves continue to depart from the real data, overlapping
27\% and 40\% of observations, respectively (fewest of all models).
The bounce at long wavelengths is the smallest when compared to models
M1--M3, yet it is still present.

\section{Conclusions}
\label{sec:concl}

Infrared emission of the protoplanetary disk \hd{}
can possibly arise from the sublimation front of dust grains,
known as the inner rim, as well as from the dust-free region interior to the rim
\citep[e.g.][]{Tannirkulam_etal_2008ApJ...677L..51T,Benisty_etal_2010A&A...511A..74B,Setterholm_etal_2018ApJ...869..164S,Varga_etal_2021A&A...647A..56V,GRAVITY_2021A&A...654A..97G}.
In this paper, our strategy was to calculate various physical models of the inner rim
in order to assess how they compare to interferometric observables.
We used radiative hydrostatic modelling \citep[following][]{Flock_etal_2016ApJ...827..144F} to derive the structure of the inner rim, we calculated synthetic images of the NIR and MIR emission,
and we compared the half-light radii $R_{\mathrm{hl}}$, fractional disk fluxes $\mathcal{F}=F_{\mathrm{disk}}/F_{\mathrm{tot}}$, and interferometric visibilities $V$ with the VLTI multi-band data \citep[e.g.][]{Lazareff_etal_2017A&A...599A..85L,GRAVITY_2019A&A...632A..53G,Varga_etal_2021A&A...647A..56V}.
Of the three quantities, $V$ is the most and $\mathcal{F}$ 
is the least robust. Interestingly, we found theoretical arguments for $R_{\mathrm{hl}}$ 
being more robust than previously thought, as discussed in Section~\ref{sec:hlr_vs_vis}.

In our set of models, we started from a nominal model (M1)
and we gradually increased the temperature of the infrared photosphere
near the inner rim by reducing the cooling efficiency of the disk (model M2),
changing the rim geometry from rounded to wall-like (model M3), and allowing the dust grains
to survive at larger temperatures (model M3Fe).
We concluded that model M3 is the one closest to observations because
it provides the best match to the visibility curves and it reproduces
previously reported $R_{\mathrm{hl}}$. It can also match $\mathcal{F}$ fairly
well, with the exception of bands H and K for which earlier morphological
fits \citep{Lazareff_etal_2017A&A...599A..85L,GRAVITY_2019A&A...632A..53G}
predict fractional disk contributions larger by a factor of 1.4 and 1.1,
respectively, compared to our model.

However, we pointed out that our model visibility curves always exhibit a bounce
at long baselines, most notably in the H band, due to the fact that the rim emits as a narrow
torus. Such bounce is not observed, confirming the need for an additional
emission component in the disk \citep[e.g.][]{Benisty_etal_2010A&A...511A..74B,Setterholm_etal_2018ApJ...869..164S}.
Additionally, matching the visibility curves across multiple bands
with a rim model alone is clearly challenging
because even though model M3 provides the best match on average, other models outperform it
when focusing on single bands (e.g. model M1 is better in the L band; model M2 is better in the K band). In other words, when one tries to modify
the physical model to improve the match in a single band, the match in other
bands might actually become worse.

In model M3, the reduced cooling efficiency was achieved by setting a warmer boundary
condition for the escape of photons by radiative diffusion and the wall-like shape 
was obtained owing to the uniform sublimation temperature of dust grains.
The realism of both model ingredients is debatable.
The reduced cooling efficiency was used in similar forms in 
the majority of recent works oriented on the inner rim \citep{Flock_etal_2016ApJ...827..144F,Schobert_etal_2019ApJ...881...56S}
but we pointed out (Appendix~\ref{sec:app})
that it leads to temperature profiles warmer than, and therefore inconsistent with,
those resulting from frequency-dependent Monte Carlo calculations.
The uniform sublimation temperature, on the other hand, used to 
be assumed in the classical inner rim models \citep[e.g.][]{Dullemond_etal_2001ApJ...560..957D}
but was later abandoned since the local conditions in terms of vapor densities
at saturation pressures are expected to change with height above the disk midplane \citep{Isella_Natta_2005A&A...438..899I,Kama_etal_2009A&A...506.1199K}.
 Nevertheless, even if the reduced cooling efficiency or the uniform sublimation temperature
 turn out to be inadequate, they might still point to the correct
disk structure, which future models could strive to reproduce by considering additional physical processes, for instance, thermo-chemical effects or high-energy non-equilibrium heating of the gas atmosphere.

To summarize physical features of model M3, the rim extends between 0.3 and 0.47 au,
with $R_{\mathrm{hl}}\simeq(0.28,0.29,0.31,0.97)\,\mathrm{au}$ and
$\mathcal{F}\simeq(0.32,0.75,0.94,1.0)$
at $\lambda=(1.5,2.2,3.5,10.5)\,\mu\mathrm{m}$, respectively.

The impact and future applicability of our study can be threefold. 
First, while parametric morphological fitting is by far the most common approach to interpret
sparse interferometric observations of sub-au disk regions, it is rarely done 
in a multi-band manner and if so, it often lacks a link to physical models.
Therefore, synthetic data from physical models could be used to calibrate morphological fits
(by checking if the fit can retrieve the important features of the physical model) before applying those to real observations.
Second, the emission component interior to the rim in \hd{} has not yet been described with a physical model.
When such a description becomes available, our study can help
tweaking the relative contribution between the rim and the additional interior source to the overall signal.
Third, radiative hydrostatic models of the inner rim are notoriously known for not producing enough NIR flux
\citep{Vinkovic_etal_2006ApJ...636..348V,Turner_etal_2014ApJ...780...42T,Flock_etal_2016ApJ...827..144F}
and our findings provide ways how to boost the flux when needed (note that model M3Fe
matches the observed $\mathcal{F}$ very well).

\begin{acknowledgments}
Based on observations collected at the European Southern Observatory under ESO programme(s) 0101.C-0896(D), 0103.C-0347(A,C), 0103.C-0915(A,B,C), 0103.D-0153(C,D), and 0103.D-0294(A).
This work was supported by
the Czech Science Foundation (grant 21-23067M),
the Charles University Research Centre program (No. UNCE/24/SCI/005)
and the Ministry of Education, Youth and Sports of
the Czech Rep. through the e-INFRA CZ (ID:90254).
M.F. acknowledges funding from the European Research Council (ERC) under the European Union’s Horizon 2020 research and innovation program (grant agreement No. 757957).
T.U. acknowledges the support of the DFG-Grant ``INSIDE: The INner regions of protoplanetary disks:SImulations anD obsErvations'' (project number 465962023).
M.B. has received funding from the European Research Council (ERC) under the European Union's Horizon 2020 research and innovation programme (grant agreement No. 101002188).
We sincerely thank Kees Dullemond and Roy van Boekel for motivating discussions,
and Jozsef Varga for sharing the data 
from \cite{Varga_etal_2021A&A...647A..56V}
with us.
We wish to thank two anonymous referees whose valuable
comments allowed us to greatly improve this paper.
\end{acknowledgments}

%

\vspace{5mm}
\facility{VLTI(PIONIER, GRAVITY, MATISSE)}


\software{\textsc{Astropy} \citep{astropy:2013,astropy:2018},
          \textsc{Fargo3D} \citep{Benitez-Llambay_Masset_2016ApJS..223...11B},
          \textsc{Matplotlib} \citep{Hunter_2007_Matplotlib},
          \textsc{NumPy} \citep{Harris_etal_2020_Numpy},         
          \textsc{optool} \citep{Dominik_etal_optool_2021ascl.soft04010D},
          \textsc{pmoired} \citep{Merand_2022SPIE12183E..1NM},
          \textsc{radiation\_code} \citep{Schobert_etal_2019ApJ...881...56S},
          \textsc{Radmc-3D} \citep{Dullemond_etal_2012ascl.soft02015D},
          \textsc{radmc3dPy}  (\url{https://www.ita.uni-heidelberg.de/~dullemond/software/radmc-3d/manual_rmcpy})
          }
          



\appendix

\section{Testing boundary conditions for the radiation energy density}
\label{sec:app}

The aim of this appendix is to exemplify the influence of boundary 
conditions for the radiation energy density (Section~\ref{sec:models})
on the temperature profile
of the inner disk, as well as to provide a link to previous studies.
To do so, we followed \cite{Flock_etal_2016ApJ...827..144F} and tried to 
reproduce their model designated S100, in which the surface density of 
the gas is non-evolving and uniform, $\Sigma(r)=100\,\mathrm{g}\,\mathrm{cm}^{-2}$.

\begin{figure}
    \centering
    \includegraphics[width=0.5\columnwidth]{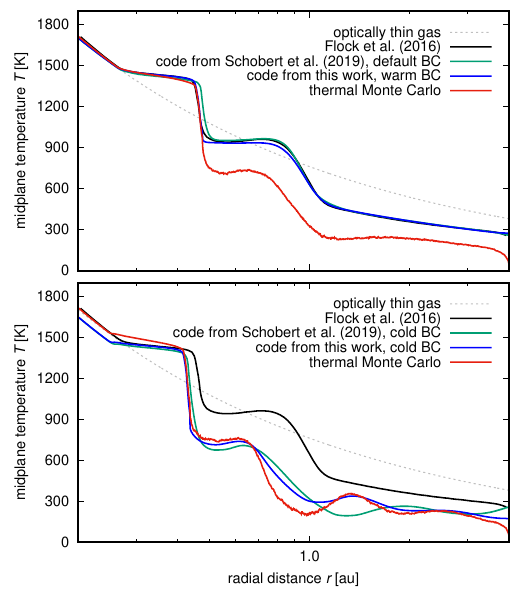}
    \caption{Radial profiles of the midplane temperature
    based on the model S100 of \citep{Flock_etal_2016ApJ...827..144F}.
    The individual panels demonstrate the influence of the boundary
    conditions for the radiation energy density (Section~\ref{sec:models})---the warm boundary is used
    in the top panel while the cold boundary is used in the bottom panel.
    Individual curves are labelled in the plot legend
    and described in detail in Appendix~\ref{sec:app}.
    To guide the eye, we advise the reader to mainly follow
    the changes of the blue and green curves (black and grey curves remain fixed, while the red curve changes only slightly).
    }
    \label{fig:s100}
\end{figure}

The results are summarized in Figure~\ref{fig:s100}.
Let us first explain the meaning of individual curves. 
The gray dashed curves show the optically thin temperature of gas 
for reference, $T_{\mathrm{thin}}=\sqrt{R_{\star}/(2r)}T_{\star}$.
The black curves show the original result of \cite{Flock_etal_2016ApJ...827..144F}.
The blue curves are the results of calculations with the code
used in this study; the top panel corresponds to the warm boundary,
while the bottom panel shows the result for the cold boundary.
In a similar fashion, the green curves show the results of calculations
that we performed using the \textsc{radiation\_code} of \cite{Schobert_etal_2019ApJ...881...56S}, the aim being to provide
an independent sanity check. Finally, the red curves show 
temperature profiles obtained with thermal frequency-dependent
Monte Carlo simulations using \textsc{Radmc-3D}, in which the input gas and dust densities
were taken from our hydrostatic calculations (those represented by blue curves).

Ideally, all solid curves should overlap. However, we see that this is only
true in the innermost dust-free disk and in the adjacent dusty halo.
The midplane temperature of the optically thick regions ($r\gtrsim0.45\,\mathrm{au}$) exhibits differences. When the boundary condition is warm
\citep[or see][for their default boundary condition]{Schobert_etal_2019ApJ...881...56S},
the temperature profiles based on our hydrostatic calculations match that of \cite{Flock_etal_2016ApJ...827..144F}.
However, the thermal Monte Carlo calculation with \textsc{Radmc-3D}
leads to substantially lower temperatures across the disk rim and the outer disk, even though the gas and dust density is directly adopted from the hydrostatic calculations.

If, on the other hand, the cold boundary condition is used, there is an agreement between our hydrostatic calculations and the thermal Monte Carlo
run, but all these temperature profiles depart from that of \cite{Flock_etal_2016ApJ...827..144F}. 
For completeness, we point out that the waves at $r\gtrsim1\,\mathrm{au}$
in the bottom panel of Figure~\ref{fig:s100} are manifestations
of the irradiation instability
\citep[e.g.][]{Watanabe_Lin_2008ApJ...672.1183W,Wu_Lithwick_2021ApJ...923..123W,Melon-Fuksman_Klahr_2022ApJ...936...16M}.

It is difficult to assess which of these boundary conditions is more realistic. In general, it seems that the cold boundary is more
common in models with radiative diffusion \citep[e.g.][]{Bitsch_etal_2013A&A...549A.124B,Kolb_etal_2013A&A...559A..80K}
and it also leads to a better match with the Monte Carlo multi-frequency
approach, which is physically superior to the simple radiative diffusion.
However, it is still instructive not to disregard the warm boundary
because, as shown in the main text, it can sometimes lead to a better match
with observations, thus laying valuable groundwork for future studies.

\section{On the role of viscous heating and convergence of the hydrostatic method}
\label{sec:app_convergence}

\begin{figure}
    \centering
    \includegraphics[width=0.51\columnwidth]{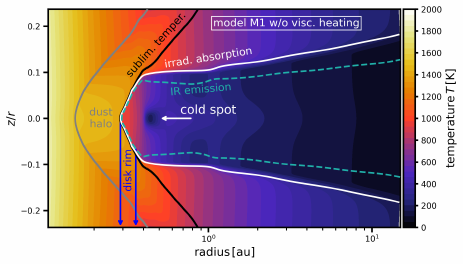}
    \includegraphics[width=0.47\columnwidth]{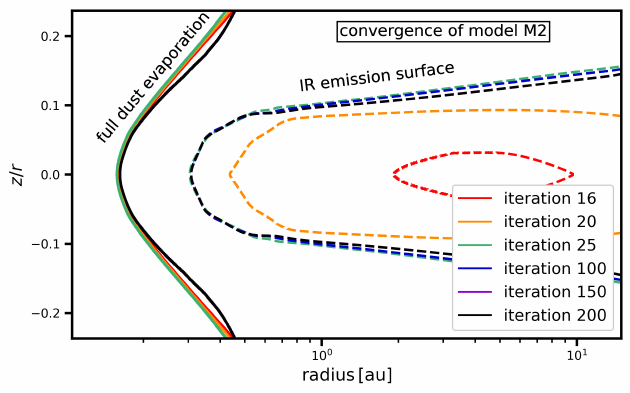}
    \caption{Left: Temperature map of model M1 with neglected viscous heating.
    There is a shadowed region outwards from the rim and an apparent cold spot. 
    Right: Evolution of surfaces of full dust evaporation (solid curve)
    and IR emission (dashed curve) during the full iteration cycle of model M2.
    }
    \label{fig:converg}
\end{figure}

Models presented in our work include a simplified treatment for viscous heating,
as explained in Section~\ref{sec:iter_4_temper}. In terms of optically thick regions
of the disk, viscous heating is implemented correctly in the dead zone of the disk
but its influence is underestimated at the very tip of the rim (because we use $\alpha_{\mathrm{DZ}}$ to evaluate the viscous heating term).
Nevertheless, it plays an important role for the stability and convergence of our models,
especially those employing the cold boundary condition for $E_{\mathrm{R}}$
(see Section~\ref{sec:models}).
To illustrate that, the left panel of Figure~\ref{fig:converg} shows a variation
of model M1 computed without viscous heating, meaning that the disk is only passively
irradiated. Without viscous heating, the disk region outwards from the rim
falls into a shadow and progressively becomes colder and colder. The reason is that
at one point during the iterative sequence, the surface density outwards from the rim
becomes so large that the vertical cooling starts to act more efficiently than heating by
radial radiative diffusion. Since no other heat source is operating in this region (due to the shadowing by the rim), the temperature slightly decreases, which means that the local viscosity $\nu$ (see Section~\ref{sec:iter_1_density}) decreases as well. Because we impose constant $\dot{M}$
through the disk, Equation~(\ref{eq:mdot}) dictates that the local surface density $\Sigma$
has to increase to compensate for lower $\nu$, leading to a feedback loop that creates a cold spot visible
in Figure~\ref{fig:converg}. Models that `fall' into this loop cannot be
reliably converged and it is questionable whether they are physically realistic (we expect that
the local density peak $\Sigma\sim3\times10^{4}\,\mathrm{g}\,\mathrm{cm}^{-2}$ overlapping with the cold spot would become Rossby-unstable
in a hydrodynamic run).
Viscous heating, even in our simplified form,
helps circumventing such convergence issues.

Our models discussed in the main body of the paper are well converged, reaching
the relative change in the temperature during last iterations of the order of $10^{-5}$.
The right panel of Figure~\ref{fig:converg} shows the evolution of several characteristic 
surfaces in model M2 during selected iterations (iteration number 200 is the last one). 
We can see that the front of full dust evaporation, which is the inner boundary of the dusty
halo, is converged already after $\sim$100 iterations since the solid blue curve is hidden underneath
the solid black curve. The IR photosphere is converged after $\sim$150 iterations because the dashed purple
curve is indistinguishable from the dashed black one (the relative difference between the two curves is below $10^{-3}$).

Finally, let us verify our assumptions
concerning the characteristic timescales in the disk (Section~\ref{sec:hydrost_model}). 
We focus on model M2 and the characteristic radial distance $r_{\mathrm{c}}=0.4\,\mathrm{au}$,
roughly corresponding to the middle of the rim extent. The dynamical timescale is 
$t_{\mathrm{dyn}}=1/\Omega_{\mathrm{K}}(r_{\mathrm{c}})\simeq 10^{6}\,\mathrm{s}$
and the viscous timescale is $t_{\mathrm{vis}}=1/(\alpha_{\mathrm{MRI}}h^{2}\Omega_{\mathrm{K}}(r_{\mathrm{c}}))\simeq 10^{10}\,\mathrm{s}$ (with the local aspect ratio $h=0.027$).
The timescale of thermal relaxation is determined by the radiative diffusion in the vertical direction from the midplane
towards the infrared photosphere. The thermal diffusivity due to radiation in the optically thick limit
is \citep[e.g.][]{Lin_Youdin_2015ApJ...811...17L,Jimenez_Masset_2017MNRAS.471.4917J}
\begin{equation}
    \chi_{\mathrm{t}} = \frac{16\sigma T^{3}}{3\rho^{2}c_{V}\kappa_{\mathrm{d}}(T_{\mathrm{s}})f_{\mathrm{d2g,max}}} \, ,
    \label{eq:diffusivity}
\end{equation}
leading to $\chi_{\mathrm{t}}\simeq1.7\times10^{16}\,\mathrm{cm}^{2}\,\mathrm{s}^{-1}$ for $T\simeq860\,\mathrm{K}$ and $\rho\simeq10^{-9}\,\mathrm{g}\,\mathrm{cm}^{-3}$. Taking the local height of the infrared photosphere $H_{\mathrm{IR}}(r_{\mathrm{c}})\simeq2.4H \simeq 0.026\,\mathrm{au}$, we obtain $t_{\mathrm{rad}}=H_{\mathrm{IR}}^{2}/\chi_{\mathrm{t}}\simeq10^{7}\,\mathrm{s}$.
Our estimates yield the inequality $t_{\mathrm{dyn}} < t_{\mathrm{rad}}\ll t_{\mathrm{vis}}$, which is consistent with \cite{Flock_etal_2016ApJ...827..144F}.

\bibliography{references}{}
\bibliographystyle{aasjournal}


\end{document}